\documentclass[preprint]{aastex}

\newcommand{\kms}{ km s$^{-1}$}
\newcommand{\ergflux}{ergs s$^{-1}$ cm$^{-2}$}

\usepackage{epsfig}
\usepackage{color}
\definecolor{brown}{rgb}{0.6,0.4,0.2}
\definecolor{purple}{rgb}{0.5,0,0.5}
\slugcomment{draft: \today , printed \today }

\title{Dust processing in Supernova Remnants: {\it Spitzer} MIPS SED and IRS Observations}
\shorttitle{Dust processing in Supernova Remnants}
\shortauthors{Andersen et al.}

\begin{document}



\author{M. Andersen}
\affil{Spitzer Science Center, California Institute of Technology, Pasadena, CA 91125}
\affil{Research \& Scientific Support Department,    European Space Agency,    ESTEC,    Keplerlaan 1,    2200 AG Noordwijk,    Netherlands\\  contact: manderse@rssd.esa.int}

\author{ J. Rho, W.~T.  Reach}
\affil{SOFIA/USRA
NASA Ames Research Center
Mail Stop N211-3
Moffett Field, CA 94035}
\author{J.~W. Hewitt}
\affil{   NASA/GSFC, 
     Mail Code 662, 
     Greenbelt, MD 20771 }
\author{J.P. Bernard}
\affil{ Centre d' Étude Spatiale des Rayonnements, CNRS, 9 av. du Colonel Roche, BP 4346, 31028 Toulouse, France}



\begin{abstract}
We present {\it Spitzer}  MIPS SED and IRS observations of 14 Galactic Supernova Remnants previously identified in the GLIMPSE survey. 
We find evidence for SNR/molecular cloud interaction through detection of [O~I] emission, ionic lines,  and emission from molecular hydrogen. 
Through black-body fitting of  the MIPS SEDs we find the large grains to be  warm, 29--66 K. 
The dust emission is modeled  using the DUSTEM code and a three component dust model  composed of populations of big grains, very small grains, and polycyclic aromatic hydrocarbons. 
We find the dust to be moderately heated, typically by 30--100 times the interstellar radiation field. 
The source of the radiation is likely hydrogen recombination, where the excitation of hydrogen occurred in the shock front. 
The ratio of very small grains to big grains is found for most of the
molecular interacting SNRs to be higher  than that found  in the plane of the Milky Way, typically by  a factor of 2--3. 
We suggest that dust shattering is responsible for the relative over-abundance  of small grains, in agreement with prediction from dust destruction models. 
However, two of the SNRs are best fit with a very low abundance of carbon grains to silicate grains and with a very high radiation field. 
A likely reason for the low abundance of small carbon grains is sputtering. 
We find evidence for silicate emission at 20 $\mu$m in their SEDs, indicating that they are young SNRs based on the strong radiation field necessary to reproduce the observed SEDs.

\end{abstract}





\section{Introduction} 

Supernovae have a profound effect on their
environment.  Supernova remnants (SNRs) will compress and heat the
surrounding interstellar material and potentially alter its chemical
composition.  Due to the short life time of massive stars, it is
expected that a significant fraction of massive stars will explode as
supernovae in or nearby molecular clouds.  However, until recently, only
 a few SNRs are known interacting with a surrounding molecular
cloud.  

Using {\it Spitzer} GLIMPSE survey data \citep{churchwell},  \citet{reach}
identified a series of Galactic SNRs in the infrared, a
sub-sample having  IRAC colors that could indicate emission from shocked  
H$_2$.    The analysis of follow-up {\it Spitzer} IRS low resolution spectra confirmed 
the detection of H$_2$ in all of the SNRs (6 from \cite{hewitt} and the rest in this paper), more than
doubling the sample of known interacting Galactic SNRs. 

Interacting SNRs are an ideal laboratory to study the effects of fast
shocks on the interstellar material.   A series of  papers have
predicted that shocks can both sputter and shatter the interstellar dust
and thus potentially modify  its abundance and size distribution \citep[e.g.][]{borkowski,jones,guillet,guillet2}.  
Changes in the dust distribution and abundance will impact the
extinction law of the dust and SNRs are  a unique possibility to
observationally constrain the shock models. 

Previous observations of interacting SNRs are limited. 
Since they  are located in the Galactic plane, there is a
significant contribution from unrelated material along the line of
sight.   Previous studies, utilizing mainly ISO observations, of e.g. 3C391, W28, and
W44 had to rely on a fit to the general Galactic background emission to
obtain the emission from the SNR.   
ISO observations of 3C391 found
that  $\sim$1 M$_\odot$ of shock excited dust was present within the 
80 arcseconds FWHM beam \citep{reachrho}.  Conversely, the Si and
Fe lines in the spectra indicated a total amount of vaporized dust of
0.5 M$_\odot$, suggesting a dust destruction of $\sim$30\% in 3C391. 
For W28, and W44, the dust emission from the SNR was relatively weak
compared to the Galactic emission and the dust mass 
could not be constrained. 
Because of improved background subtraction which
has been possible with {\it Spitzer} data due to the spatial coverage of the slit, the additional uncertainty from the
background modeling can be reduced. 

The dust analysis of interacting SNRs using ISO data described above was limited 
to long wavelengths (greater than 40 $\mu$m).
Shorter wavelength observations are
necessary to probe the emission from the very small grains and polycyclic aromatic hydrocarbons (PAHs)
in greater details.  The ratio of the very small grain (VSG) to big grain (BG) abundances can provide
insight into  dust destruction mechanisms.  Dust destruction models
predict a different destruction efficiency for graphite and silicates, 
respectively \citep[e.g.][]{jones}.  The models further predicts a strong dependence
on grain size for the dust destruction efficiency.  Thus, a comparison
of the ratio of VSGs to BGs in the SNRs with the similar ratio
determined in the Galaxy and the LMC \citep[e.g.][]{bernard} will provide
constraints on dust destruction models. 

Here we present an analysis of the continuum and PAH emission as well as
the [O~I] 63 $\mu$m emission from a sample of inner Galaxy SNRs.   We
discuss MIPS SED observations and low resolution IRS spectra  of a sample
of 14 SNRs.  The observations were centered on the emission peaks for
each SNR identified in the GLIMPSE data \citep{reach}.  The wavelength
coverage from 5 to 80 $\mu$m ensures a good sampling of the three
main dust species and enables us to fit the continuum in greater detail
than previous mid--infrared observations that had much lower spatial resolution
and lacked proper background subtraction. 

The paper is structured as follows.  In Section 2 we present the
observations and describe the data reduction.  We outline the basic results
in Section 3. 
Modified black bodies are fitted to the long wavelength data
and used to estimate the temperature of the big grains.  We further discuss the
detection of the  [O~I] line at 63 $\mu$m in several of the MIPS SEDs.  
We fit the continuum with a three-component dust model and discuss the basic properties for each SNR in Section 4. 
In Section 5 we discuss the derived relative abundance of the dust species.  The total
amount of dust in each SNR is then estimated from the 24 and 70 $\mu$m
image by extrapolation of the results of the dust fitting to the whole
SNR.  Finally, we present our conclusions in Section 6.

\section{Observations}
We have obtained {\it Spitzer} MIPS SED observations and IRS low
resolution spectra  of a total of 14 individual Galactic SNRs as part of
cycle 3 (program ID 30585; PI: J.\ Rho).  Three of the SNRs, CTB37A, RCW103, and
3C396 were observed at two positions.  For all 17 positions we have
obtained low--resolution (LR) IRS spectra in all four sub--slits of the LR module,  covering the  5--35 $\mu$m
range as well as MIPS SEDs.  Part of the IRS data were discussed in
\citet{hewitt}. Here we present the line intensities for the remaining
SNRs and discuss the continuum emission from the whole sample.  The
basic characteristics for a subset of the SNRs discussed in this paper
were described in \citet{hewitt}. 
For the
remaining SNRs, the data  are summarized in Table~\ref{basic_info} and the details are provided in Section 4.4.
Below we describe the data reduction.

\begin{deluxetable}{ccccccccc}

\tablecaption{Basic properties of sample SNRs$^a$}

\tablehead{\colhead{Name} & \colhead{Other name} &\colhead{RA} & \colhead{DEC} & \colhead{distance} & \colhead{Diameter} 
& \colhead{Size} & \colhead{N$_H$} \\
& &\colhead{(J2000)} &\colhead{(J2000) }& \colhead{(kpc)} & \colhead{($'$)} & \colhead{(pc)} & \colhead{($10^{22}$cm$^{-2}$)}
}
\startdata
G11.2-0.3 & \nodata & 18:11:32.3 & -19:27:12  & 5.0 & 4 & 5.8 & 2.0 \\
Kes69 & G21.8-0.6 & 18:33:01.9 & -10:13:43 & 5.2 & 20 & 30 & 2.4 \\
G22.7-0.2 & \nodata & 18:33:09.0 & -09:26:41 & 3.7 & 26 & 28 & 7.8\\
3C396cent & G39.2-0.3 & 19:04:18.7 & +05:26:31 & 8.0 & 8x6 & 16 & 4.7\\
3C396shell & G39.2-0.3 & 19:03:56.2 & +05:25:50 & 8.0 & 8x6 & 16 & 4.7\\
G54.4-0.3 & \nodata & 19:33:05.5 & 19:16:47 & 3.0 & 40 & 35 & 1.0\\
Kes17 & G304.6+0.1 & 13:05:32.8 & -62:40:06 & 9.7 & 8 & 23 & 3.6\\
Kes 20A &  G310.8-0.4 & 14:00:41.4 & -62:20:21 & 13.7 & 12 & 48 & 7.9\\
G311.5-0.3 & \nodata & 14:05:22.1 & -61:58:11 & 14.8 & 5 & 21.5 & 2.5\\
RCW103 & G332.4-0.4 & 16:17:31.8 & -51:06:34 & 3.1 & 10 & 9 & 0.7\\
RCW103fill & G332.4-0.4 & 16:17:14.8 & -51:01:40 & 3.1 & 10 & 9 & 0.7\\
G344.7-0.1 & \nodata & 17:03:57.6 & -41:40:51 & 14.0 & 10 & 41 & 5.5\\
G346.6-0.2 & \nodata & 17:10:14.6 & -40:14:38 & 11 & 8 & 26 & 2.7\\
G348.5-0.0 & \nodata & 17:15:04.9 & -38:33:41 & 11.3 & 10 & 33 & 2.8\\
CTB 37A--N & G348.5+0.1 & 17:14:25.8 & -38:33:09 & 11.0 & 15 & 48 & 3.2\\
CTB 37A--S & G348.5+0.1 & 17:14:35.0 & -38:28:20 & 11.0 & 15 & 48 & 3.2\\
G349.7+0.2 & \nodata & 17:18:00.7 & -37:26:16 & 22.4 & 2.5x2 & 15 & 5.8 \\
\enddata
\tablenotetext{a}{The basic properties for the sample SNRs -- Kes 69, 3C 396, Kes 17, G346.6-0.2, G348.5-0.0, 
and G349.7+0.2 -- are obtained from  \citet{hewitt}. The references are given in the text. 
The coordinates provided are for the location of the extracted IRS spectra.}
\label{basic_info}
\end{deluxetable}

\subsection{MIPS SEDs}
Mid--infrared observations of the SNRs were obtained using the SED mode
of the MIPS instrument.  The spectral dispersion is 1.771 $\mu$m pixel$^{-1}$ over
the wavelength range 53$\mu$m to 96$\mu$m.  
Details for the MIPS
SED mode can be found in \citet{lu}.  Observations were obtained in a
standard on--off pattern with an offset of 3\arcmin .  Each cycle of
on-off positions consisted of a STIM flash and  3 subsequent on--off
pairs.  For each SNR we obtained 10 cycles with  the integration time 
chosen to be 3.15 seconds for each on and off position. We have adopted
the pipeline processed BCD frames and produced the final on and off
mosaics using the mopex software provided by the Spitzer Science Center.  However, the flash associated with the STIM was found to
contaminate the first on--off pair and these have been manually excluded
before the data were co--added.  The total integration time for the on
and off positions is 63 seconds.  The locations of the on and off
positions are superposed on the images of the SNRs shown  in
Figure ~\ref{fig_24mu}.

The location of the emission from the SNR within the slit  was in each
case identified through visual inspection and by comparison with the
MIPSGAL images at both 24 $\mu$m and 70 $\mu$m.  The spectrum was
extracted over 3 pixels in the spatial direction. The extraction width
corresponds to 29\farcs4, resulting in a total extraction area of
19\farcs6 $\times$ 29\farcs4.

We have used the same region in the slit for the on and off
position for most of the SNRs.  However, the off position was
contaminated in a few instances which has forced us to use an off
position located at a slightly different position on the chip.  For
G11.2-0.3, the orientation of the slits was such that the  off position
is contaminated by an extremely bright mid--IR source.  The source was
bright enough to contaminate the whole slit and no suitable background
could be obtained.  We have instead used a different  off position   at
(RA,DEC)=(272.618098,-19.374681) (program ID 3121, PI  K. Kraemer).  One
of the observations of RCW 103 had the off position located within the
SNR due to the particular roll angle of the satellite during the
observations.  We have used the off position from the other RCW 103
observation (RCW 103 filament) for both MIPS SEDs.  Further, we have
used the most western part of the off observation to extract the sky
since most of the off position was within the SNR as well. 

Adopting a different off position can introduce instrument artifacts that would otherwise have been cancelled out in a normal chopping pattern. 
We have used standard star observations
to investigate the potential effects due to using a different off
position.  Comparison of the different off positions show that the
typical error in the measured surface brightness is of the order 10--20
MJy/sr.  This corresponds to an additional uncertainty of 5--10\% in the
surface brightness errors.  
We find the scatter to be mainly random in a choice of a different sky position 
with the potential systematic shift being less than 10 MJy/sr. 
Thus, since the large grain dust abundance derived below is linearly dependent 
on the surface brightness, an additional systematic uncertainty of less  than 10\% 
is expected for this component in the case of G11.2-0.3. 
For RCW103 the uncertainty is $\sim$20\% for the filament position. 
The second position turns out to be very weak at long wavelengths and the continuum is thus poorly defined. 
Thus, we have only provided the surface brightness of the [O~I] line for the MIPS SED in the subsequent analysis.

The surface brightness of the Spitzer Science Center (SSC) SED pipeline
products of BCD and mosaicked images were a factor of 4-5 higher than
that of the MIPS wide-field (WF) images. 
By working with MIPS Ge team,
we found that two corrections should be made for extended emission.
First, the processed data were multiplied by the aperture correction
because they are optimized to point sources. 
The  aperture corrections as a function of wavelength is provided in Table 2 in \citet{lu}. 
Second, the MIPS images  have a  pixel size of 9.8$''$ square while the width 
of the SED slit is 20$''$. 
Thus, the BCD values should be divided
by 2.041.
More information can be found at 
the website\footnote{http://irsa.ipac.caltech.edu/data/SPITZER/docs/mips/features/}. 



The derived surface brightness have been compared with the
surface brightnesses determined from the MIPS 70$\mu$m WF images. 
The
MIPS imaging data are plagued by non--linearity effects due to the
brightness of the Galactic plane at mid--IR wavelengths.  The surface
brightness is thus consistently lower in the MIPS WF images than in the
MIPS SEDs and the flux ratio changes as the function of the flux level.  
The 'on' positions are brighter than the MIPS WF fluxed by
$\sim$ 40\%\ whereas the off positions are brighter by $\sim$30\%. 
The discrepancy is  most likely due to non linearity.  After discussions with the
MIPS SED team, we have agreed the SED surface brightness estimates are
more accurate than the MIPS WF values. 
The MIPS 160 $\mu$m data are  saturated in the Galactic plane. 


\begin{figure}
\centering
\begin{tabular}{cc}
G11.2-0.3 & Kes 69 \\
\includegraphics[width=5.6cm]{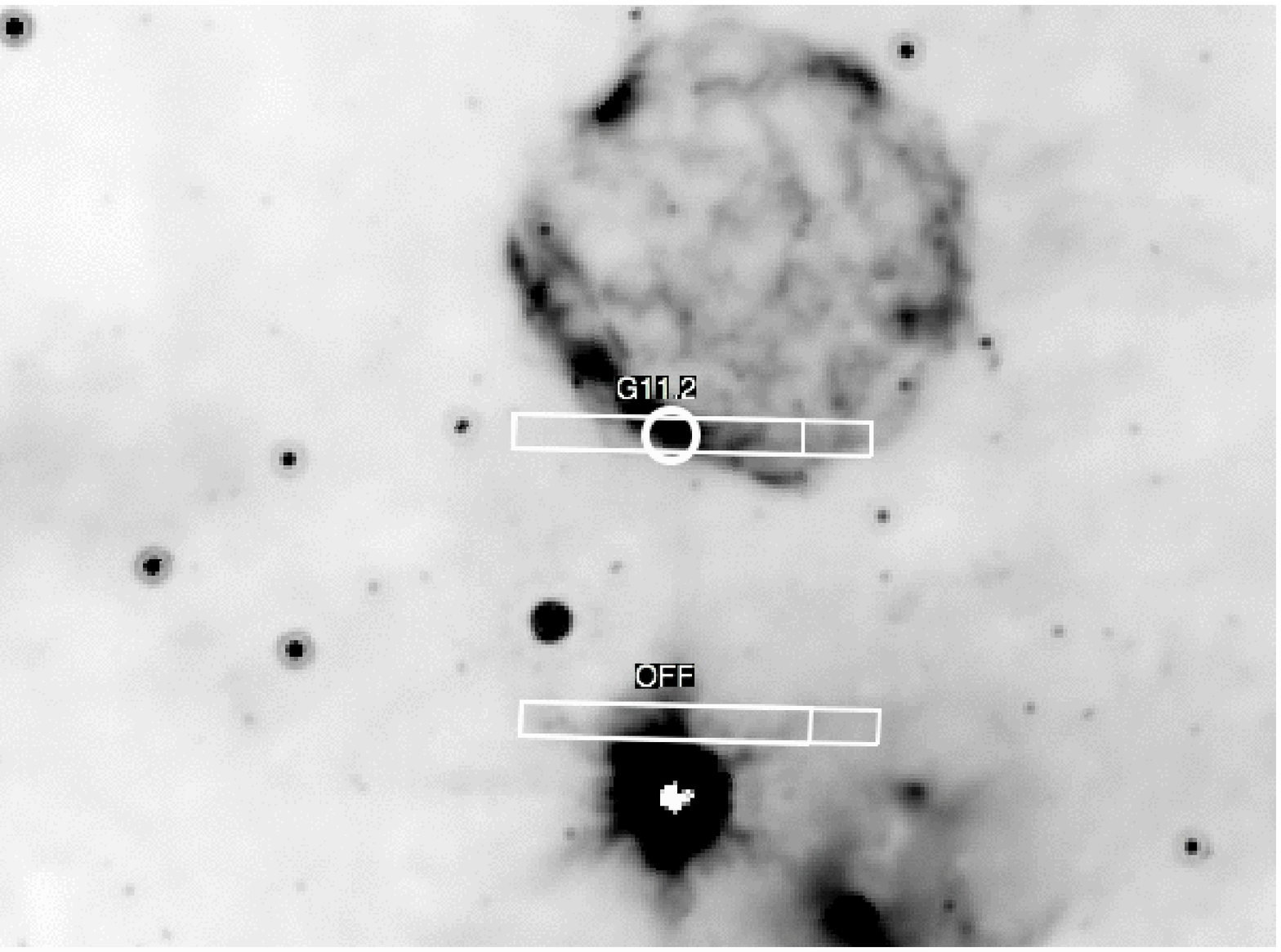}& 
\includegraphics[width=5.6cm]{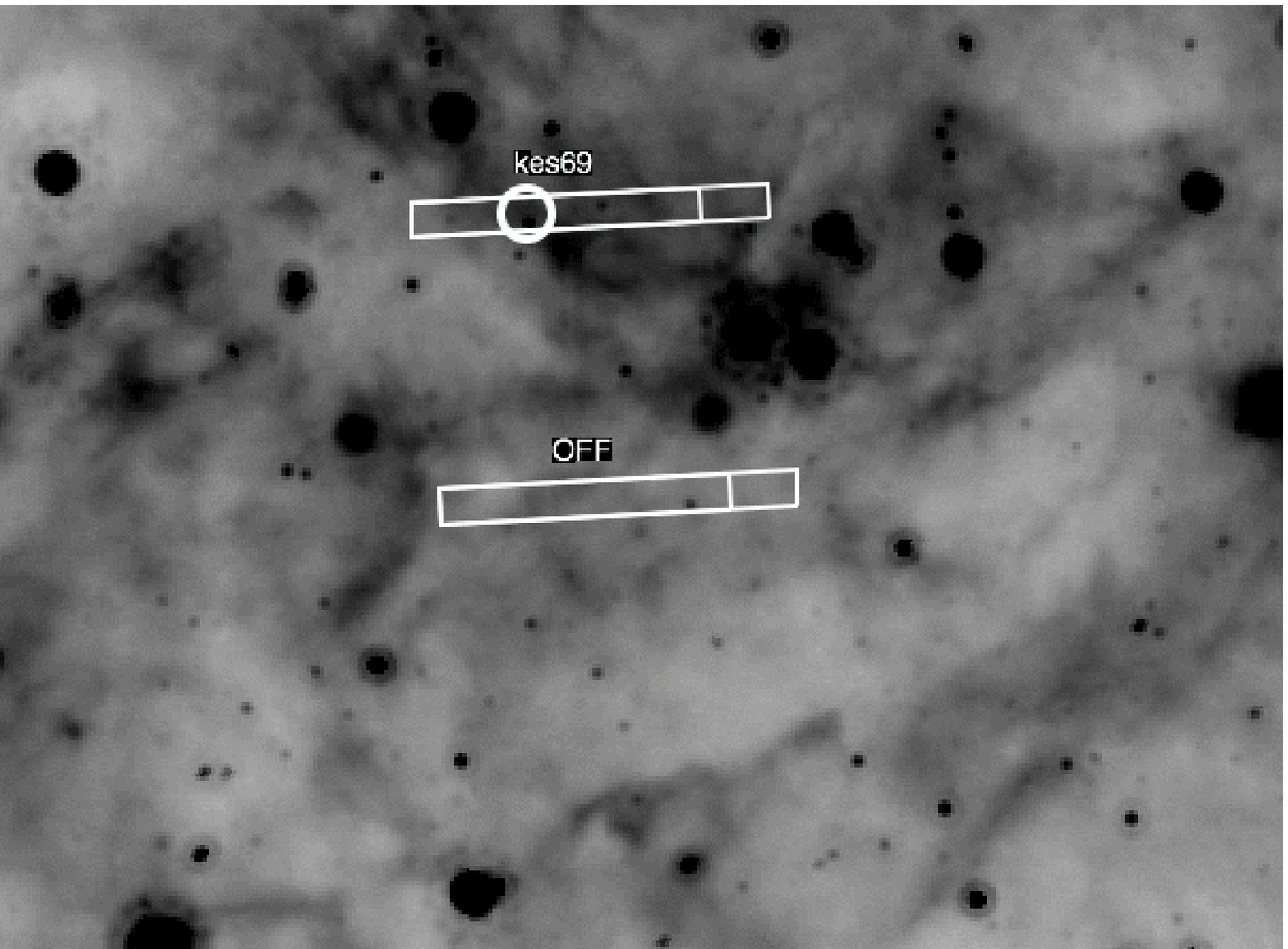} \\ 
G22.7-0.2 & 3C396 \\
\includegraphics[width=5.6cm]{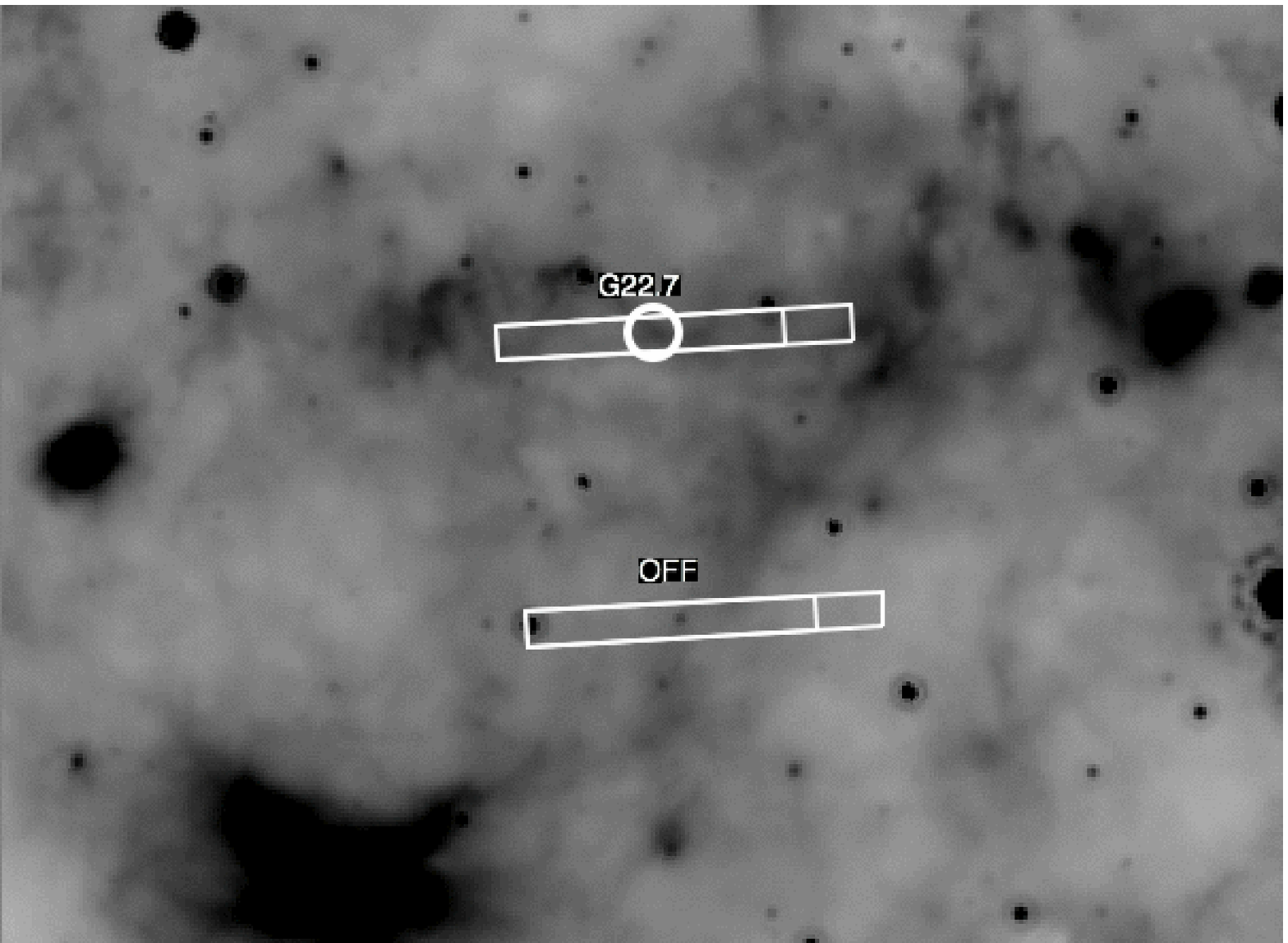} & 
\includegraphics[width=5.6cm]{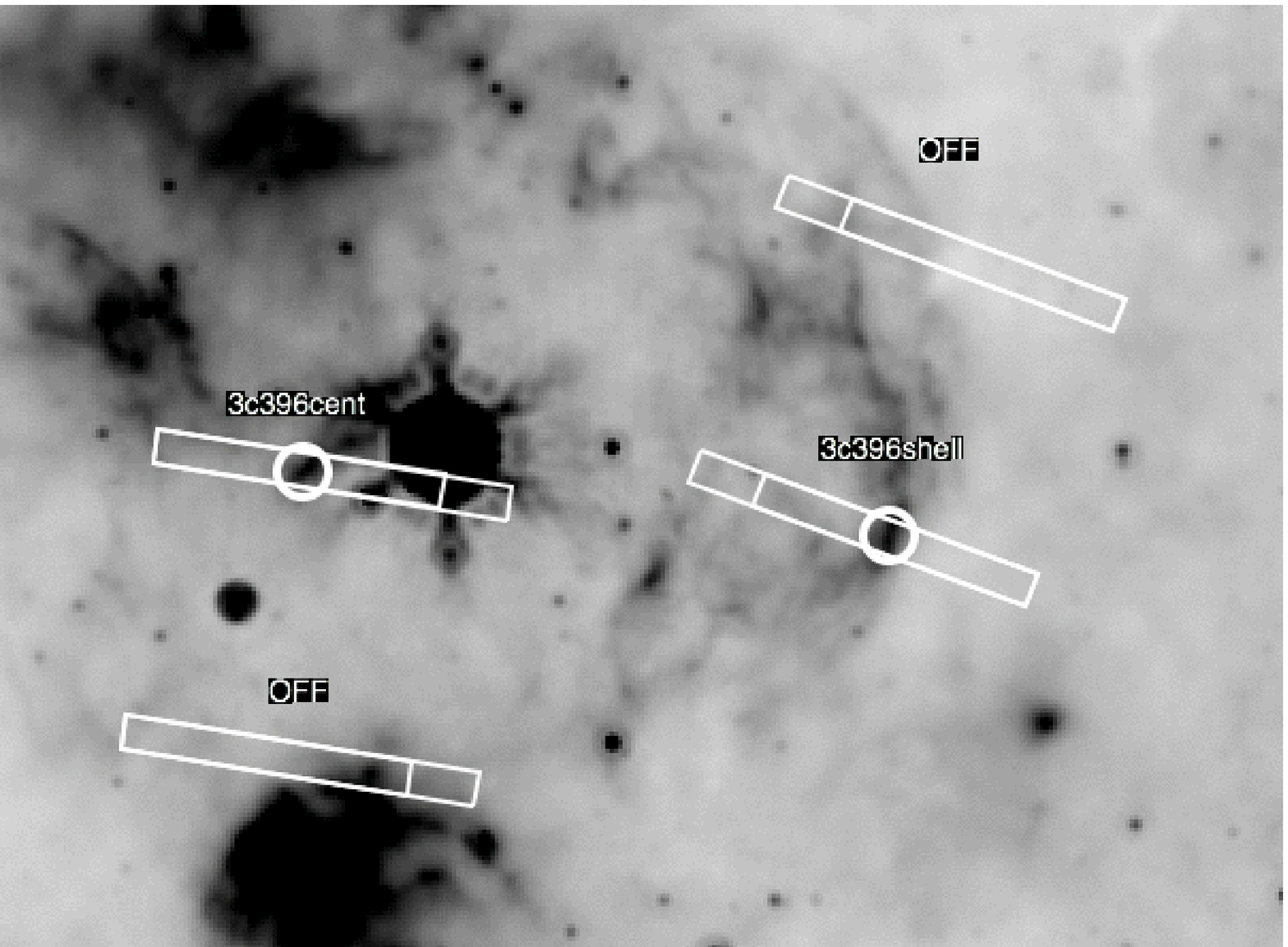} \\
G54.4-0.3 & Kes 17 \\
\includegraphics[width=5.6cm]{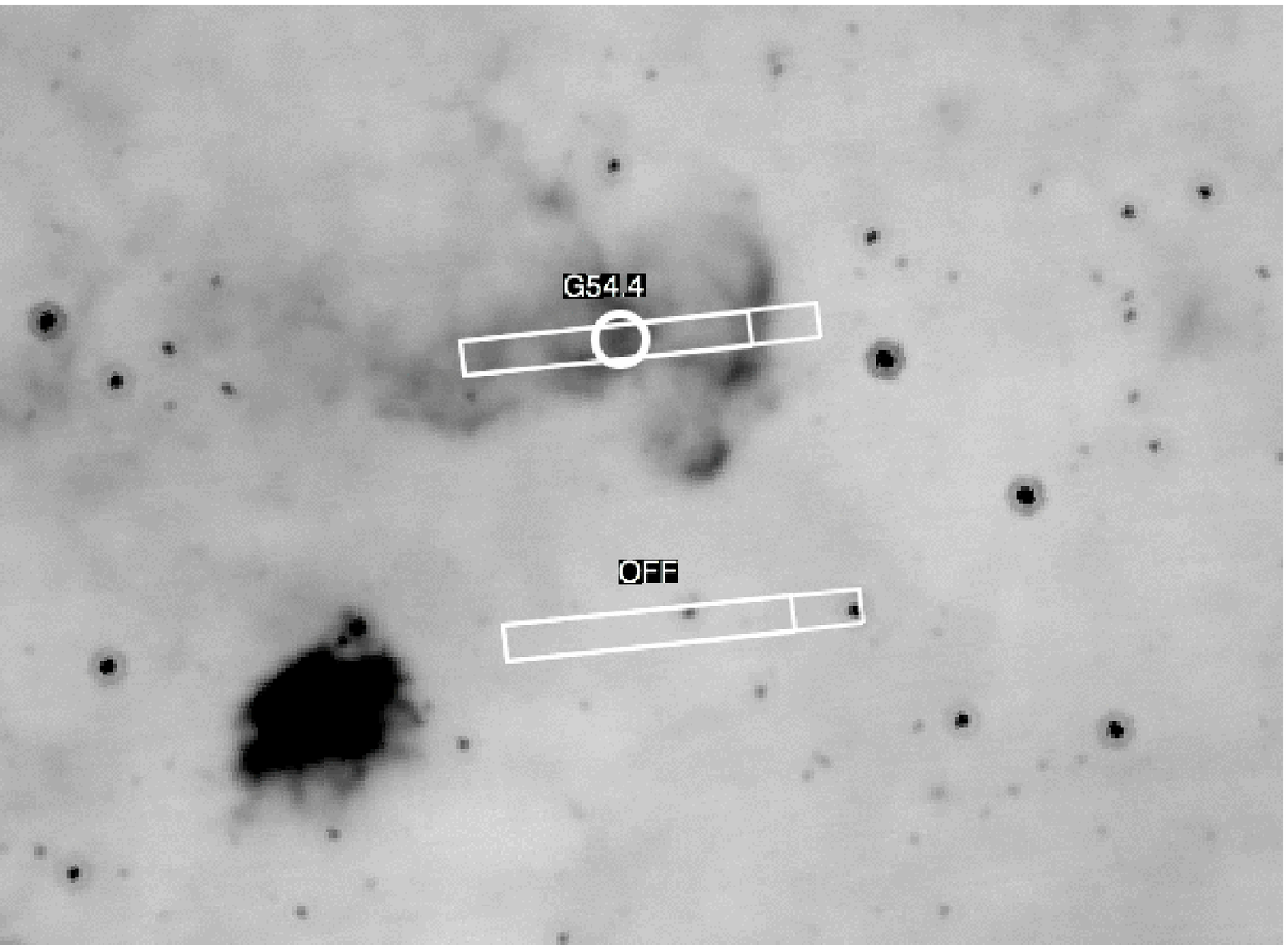} & 
\includegraphics[width=5.6cm]{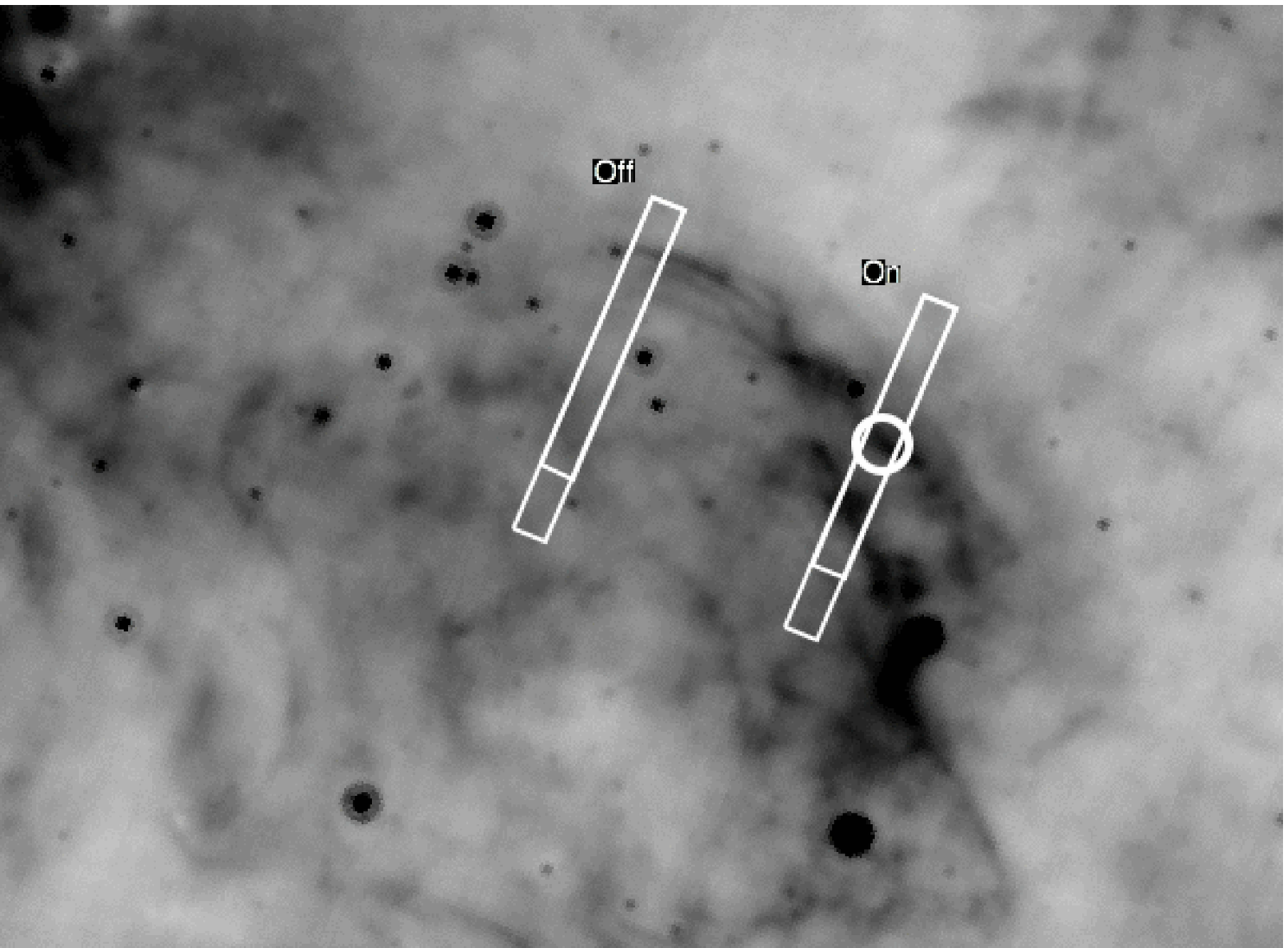} \\
Kes 20A & G311.5-0.3 \\
\includegraphics[width=5.6cm]{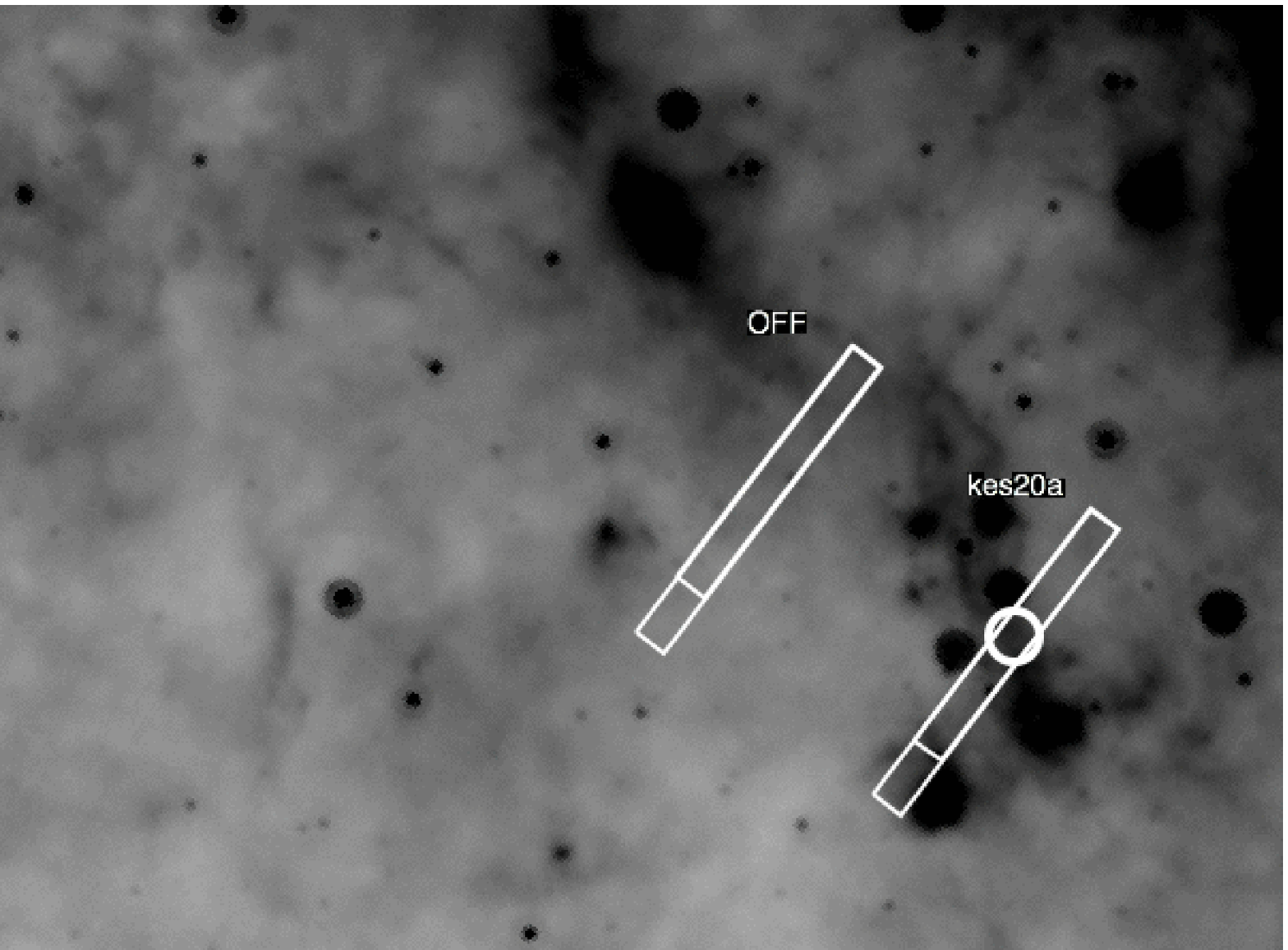} & 
\includegraphics[width=5.6cm]{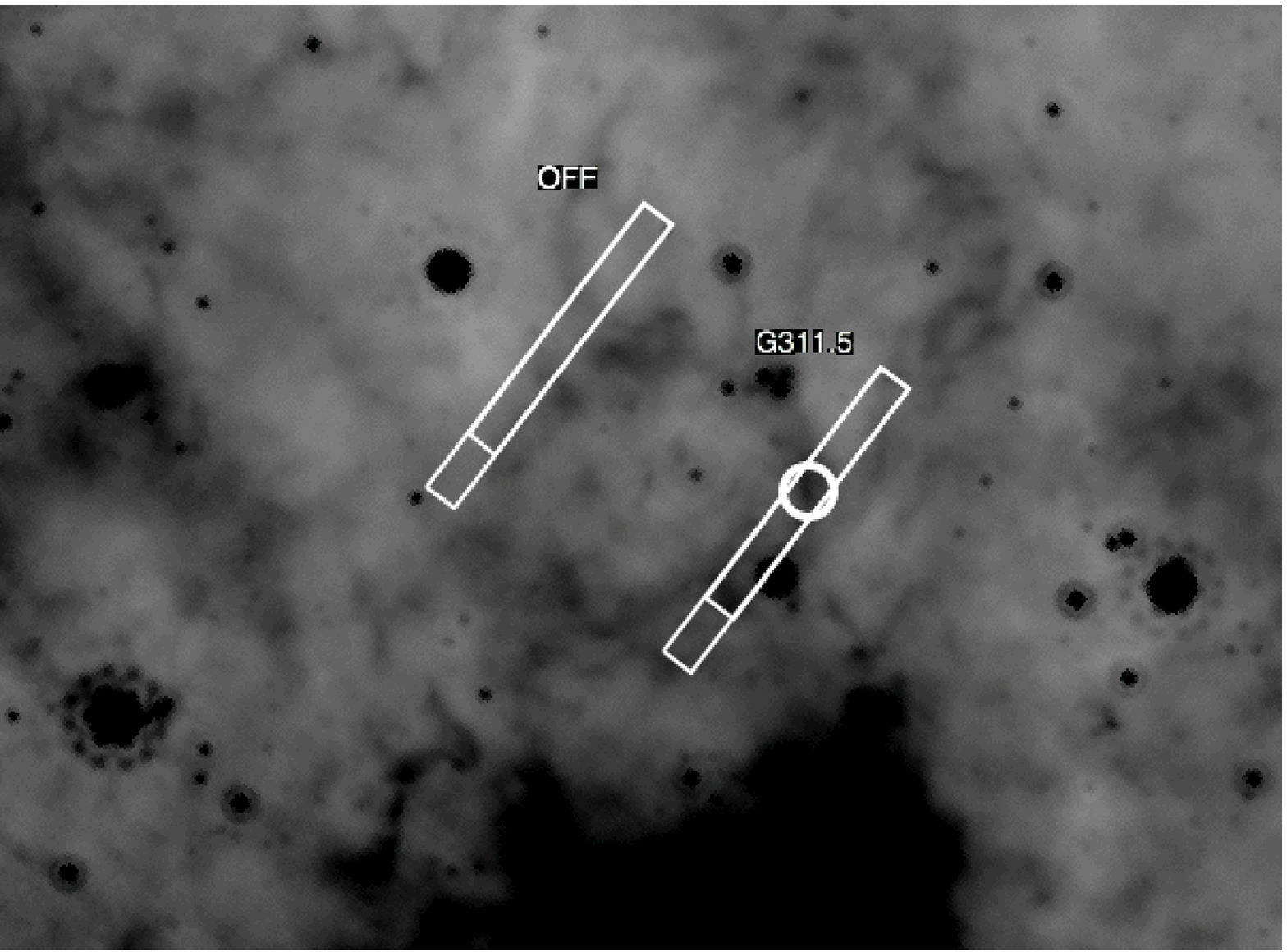} \\
\end{tabular}
\end{figure}
\begin{figure}

\begin{tabular}{cc}

RCW 103 & G344.7-0.1 \\
\includegraphics[width=5.6cm]{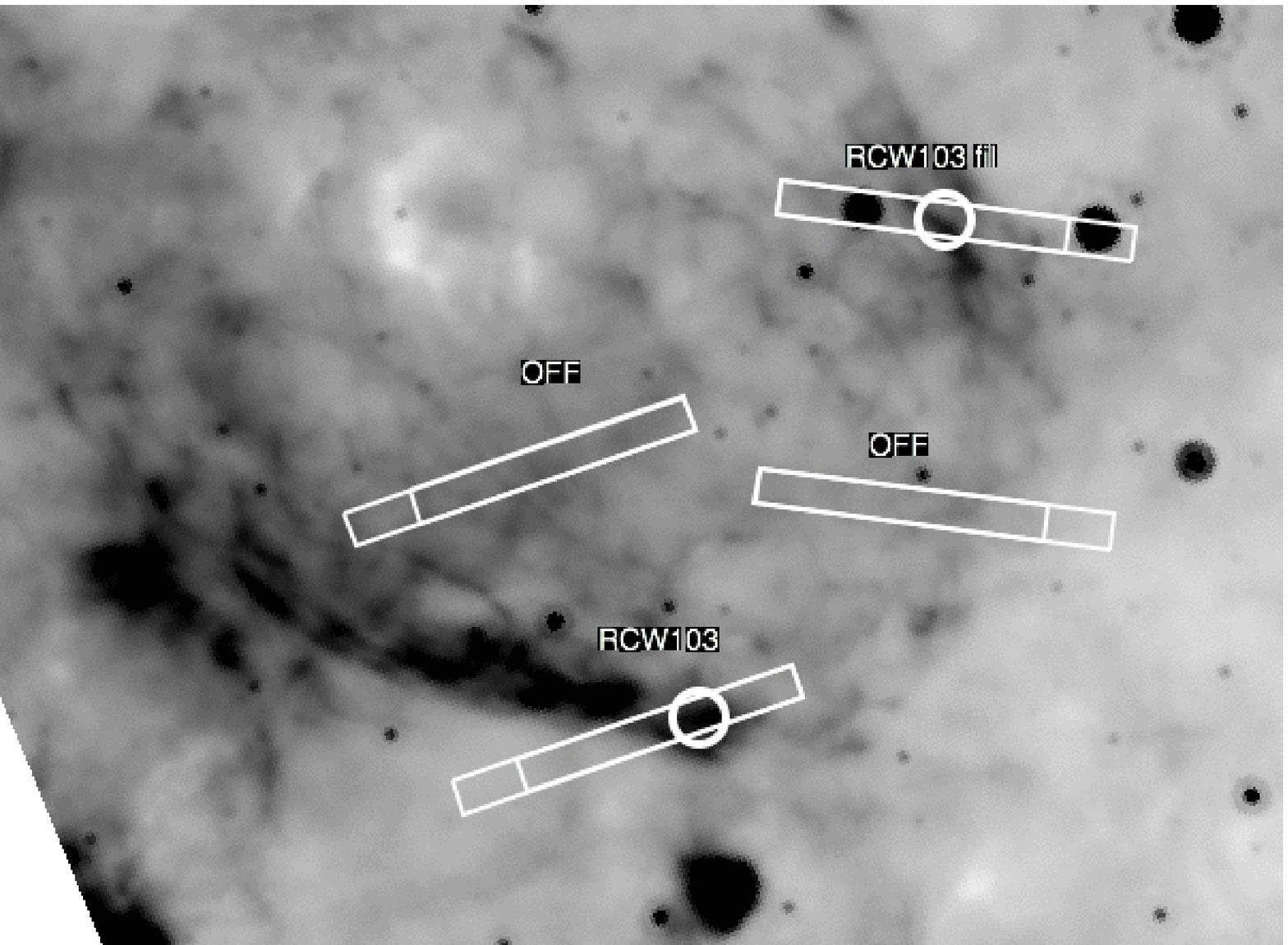} & 
\includegraphics[width=5.6cm]{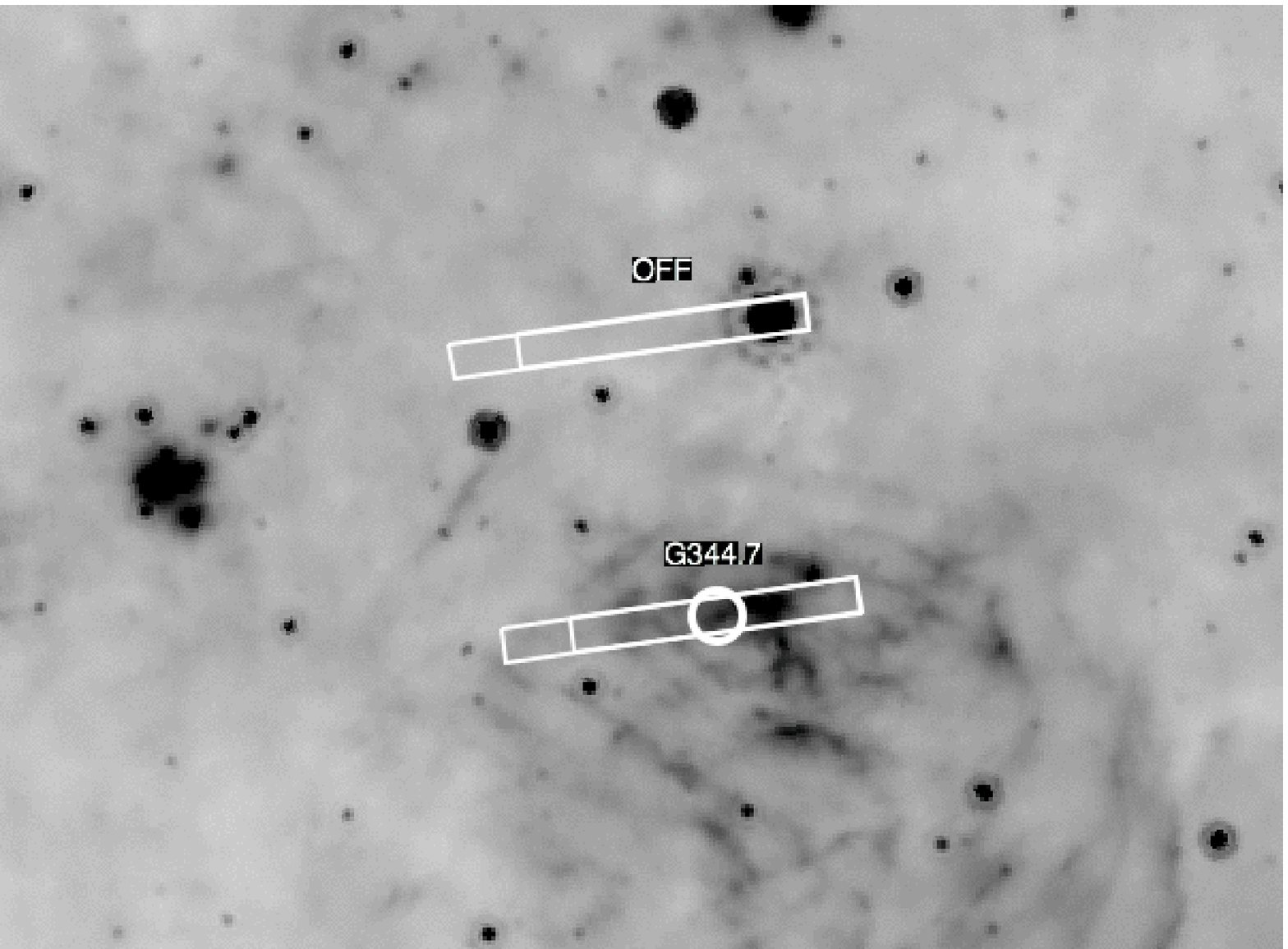} \\

G346.6-0.2 & G348.5-0.0 \\
\includegraphics[width=5.6cm]{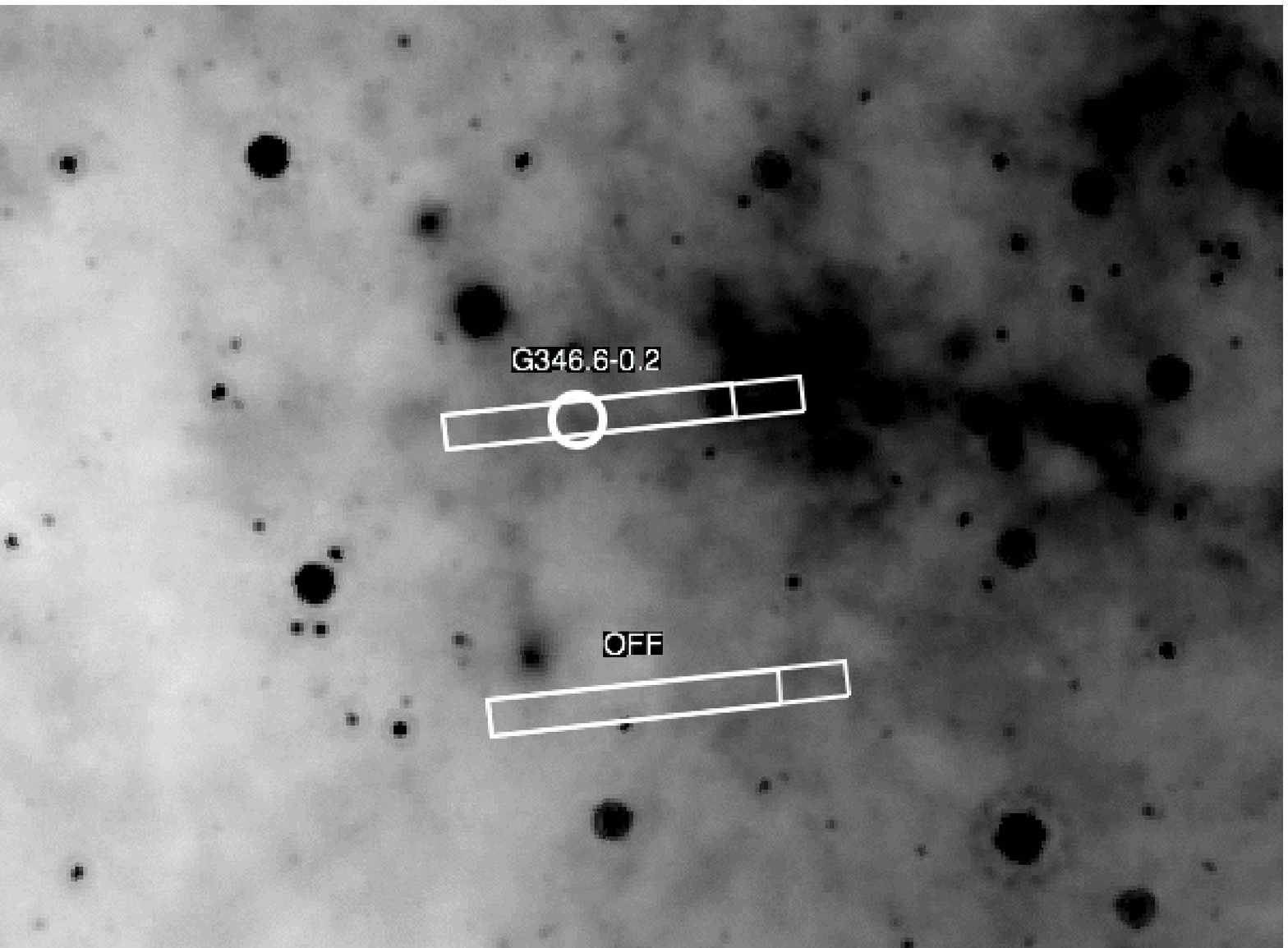} &
\includegraphics[width=5.6cm]{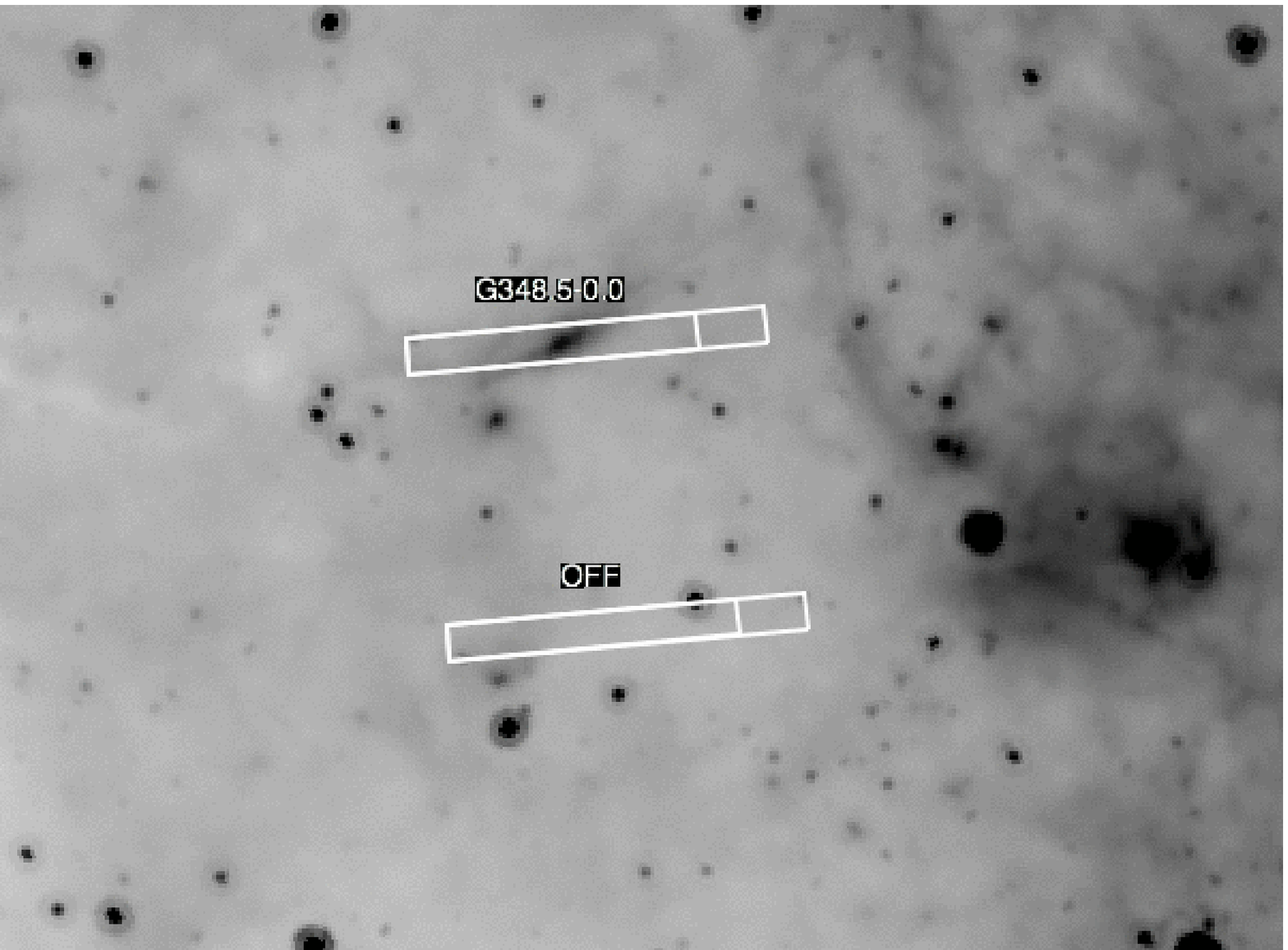} \\

CTB37A-N & CTB37A-S  \\
\includegraphics[width=5.6cm]{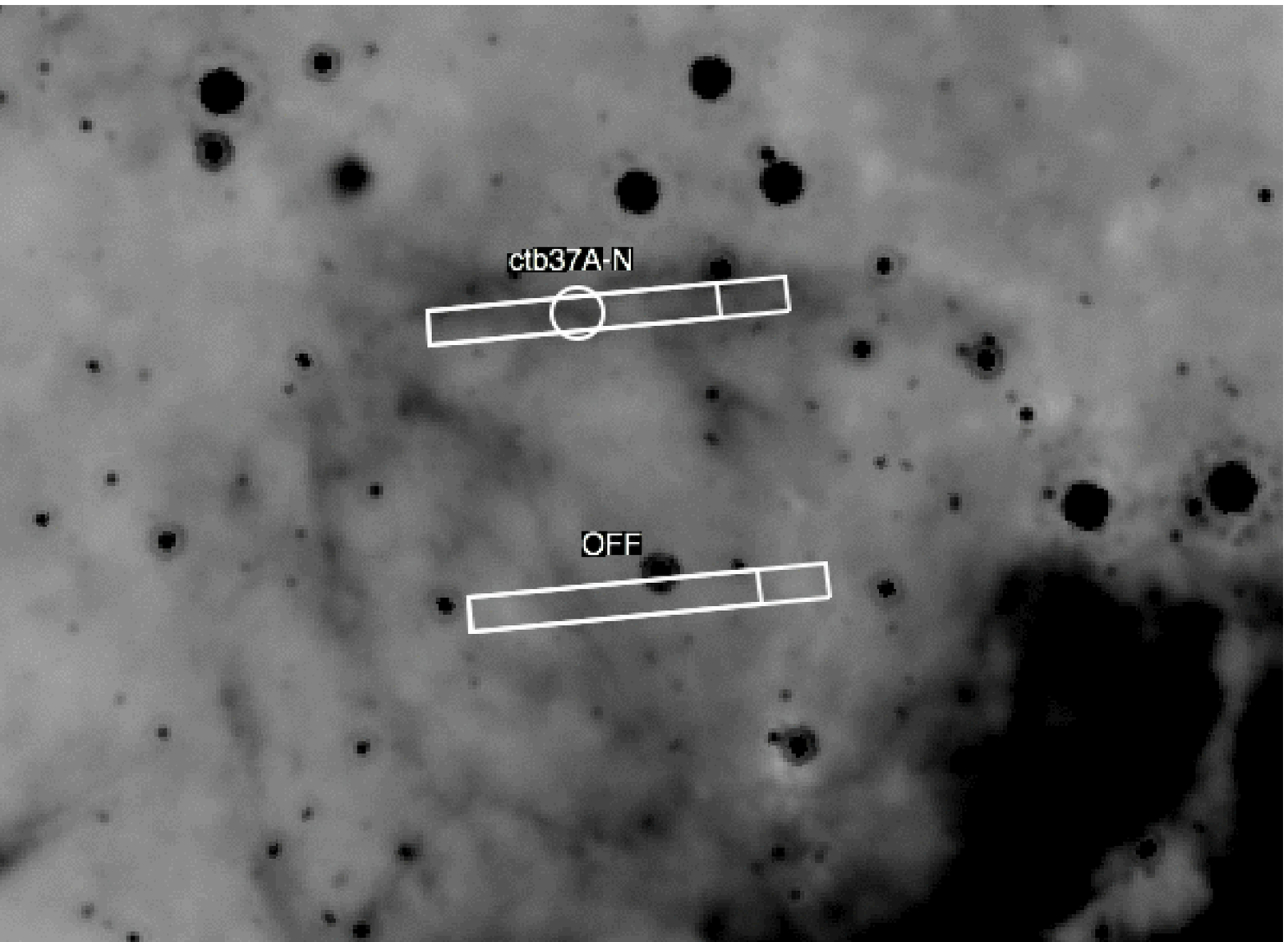} &
\includegraphics[width=5.6cm]{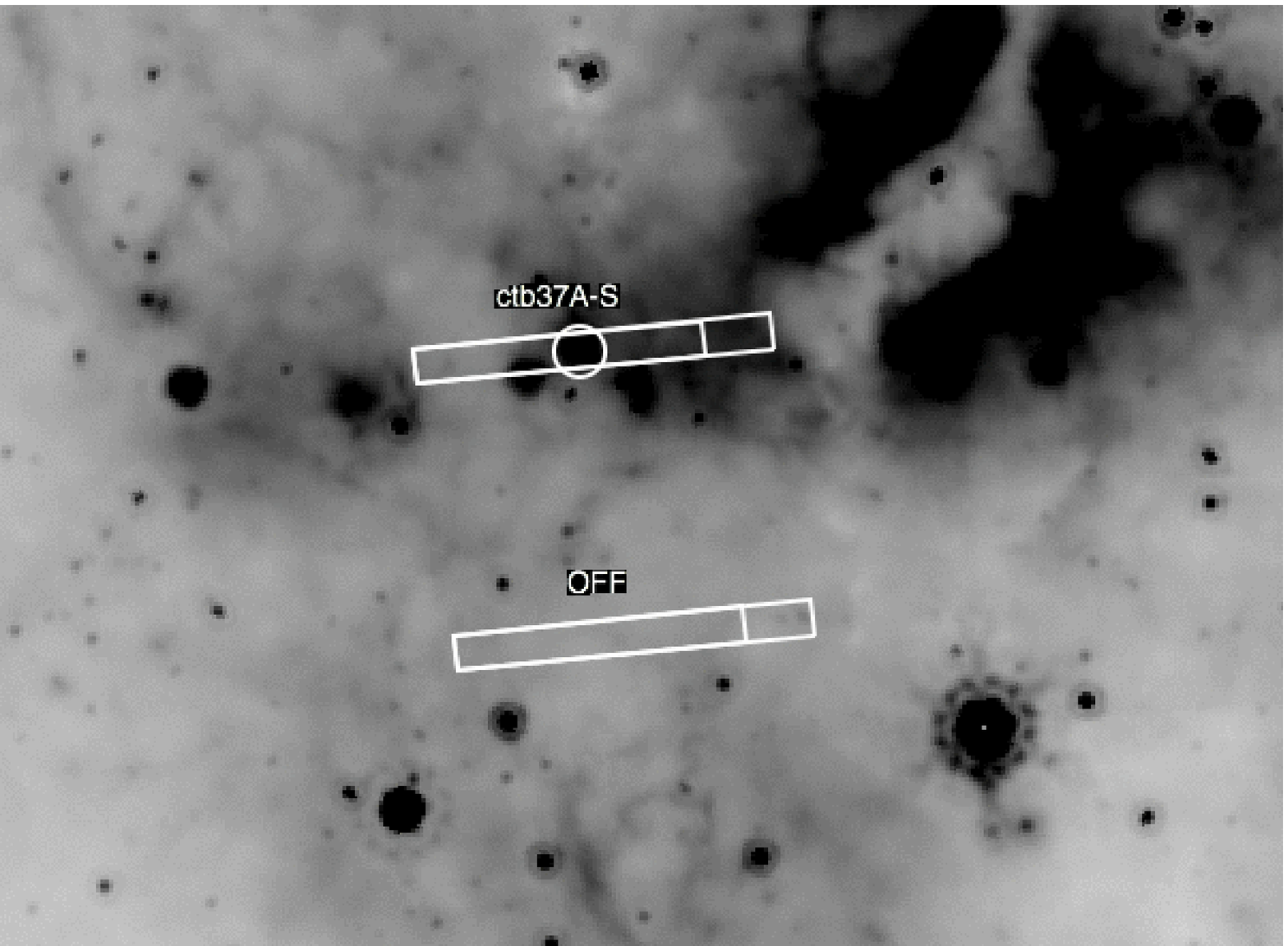} \\ 

G349.7+0.2 & \\
\includegraphics[width=5.6cm]{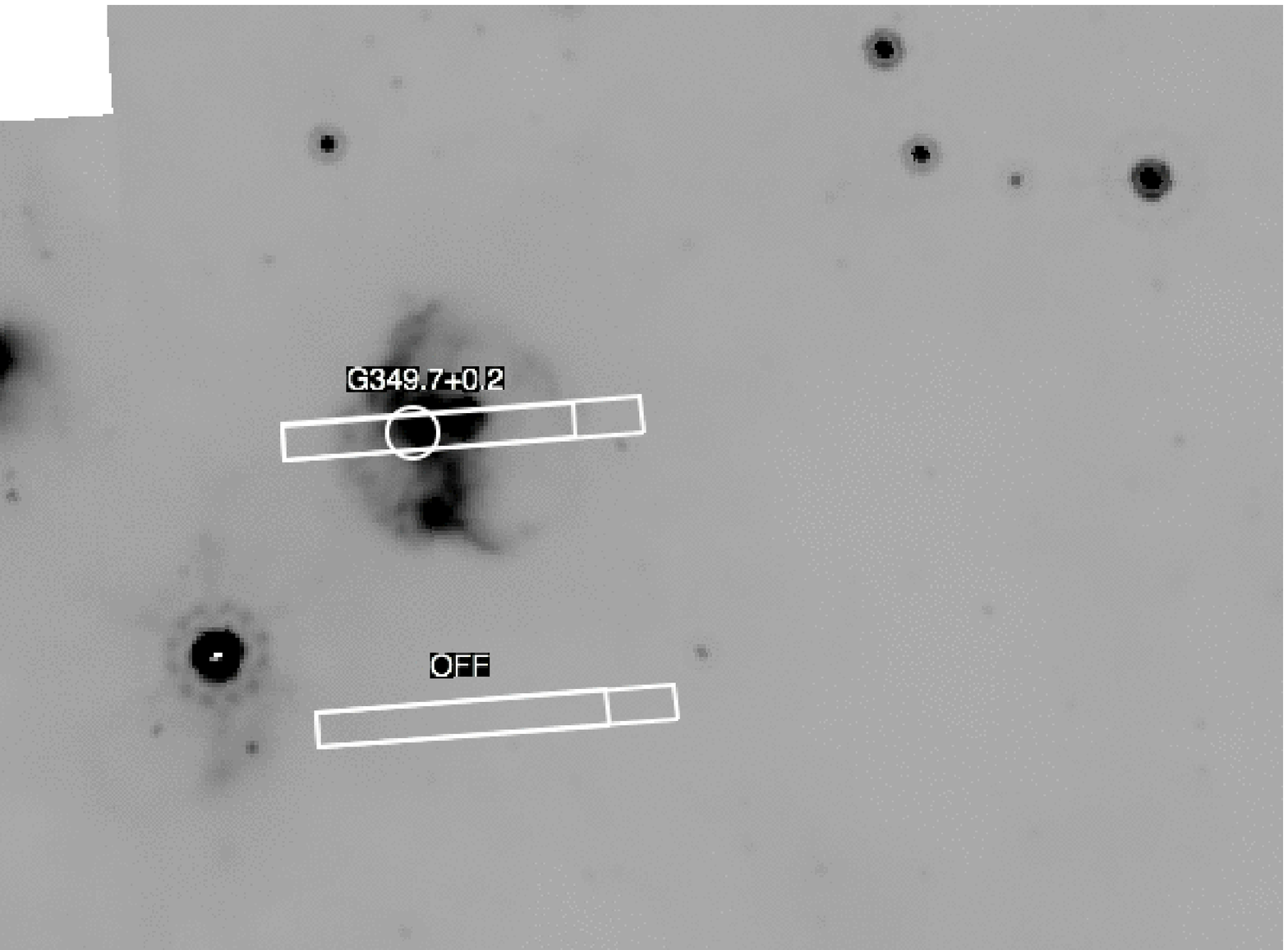} \\
\end{tabular}







\caption{  {\it Spitzer} MIPSGAL 24 $\mu$m images covering the MIPS SED slit region. 
Each image is 12\arcmin$\times$8\arcmin\ and the intensity scale is inverted. 
The location of both the on and off slits are located in each image. 
In each image the center of the circle is the central position for the extracted SEDs and IRS spectra. 
North is up, east is to the left.}
\label{fig_24mu}
\end{figure}


\subsection{IRS Spectra}
IRS low--resolution spectra were obtained for all the SNRs. 
The wavelength coverage is from 5--37 $\mu$m with a spectral resolution of 50--100.  
The 2 dimensional BCD frames were reduced in a standard manner utilizing the {\it Spitzer} pipeline. 
Adjacent frames were also obtained in order to correct the spectra for rogue pixels. 
The GLIMPSE data were used to locate regions within a few degrees of each SNR with very low emission where the IRS rogue pixel frames were obtained. 
The rogue BCD frames were median combined and were then subtracted from the BCD frames covering the SNR. 
Some rogue pixels remained and were identified by hand and replaced by the interpolated value of the surrounding pixels.

The spectrum in each sub--slit was extracted from the location of maximal emission for the SNRs, 
identified both in the long slit itself and in the IRAC/MIPS images. 
An optimal spectrum was extracted using the SPICE software. 
The central position was observed twice with each of the four sub--slits and the average spectrum was created. 
The two spectra for each sub--slit at each position agree to better than 10\%. 
Some of the spectra were reduced using CUBISM \citep{smith} which was helpful particularly 
when the sky background structures are confused with the SNRs.

There is in general substantial emission from unrelated Galactic
material along the line--of--sight of the SNRs.  The long slit nature of
our observations was used to subtract a local background from the
spectra of the SNR.  
Due to the simultaneous observations of two orders,
at two different spatial positions, we have been able to identify clean
sky positions within a few arcminutes for the majority of the SNRs. 
Since the background is subtracted from slightly different positions on
the sky, there are small differences in their surface brightness
levels.  We have corrected the individual background orders by a small
multiplicative factor ($<$10$\%$)  to ensure the spectra join in a
continuous manner. 


\subsection{MIPS Imaging}
We have further included MIPS 24 and 70 $\mu$m imaging of all the SNRs. 
The data were obtained as part of the MIPSGAL survey \citep{carey}.  

The individual BCD files covering each SNR region were obtained  from the archive,  reduced and
mosaicked  using the MOPEX pipeline.  The 24 $\mu$m images are
shown in Figure ~\ref{fig_24mu} together with the location of the MIPS SED
slits.  The morphology between the 24 and 70 $\mu$m images is typically
very similar,  but the SNRs are more
easily discernible in the 24 $\mu$m compared to the 70 $\mu$m images due to the temperature difference between 
the heated dust and the surrounding interstellar material and the artifacts present in the long wavelength data.



\section{Results}
We present the IRS spectra and MIPS SEDs below and discuss the detection
of emission from  molecular hydrogen and neutral oxygen.  The luminosity
from the dust emission is calculated for the SNRs integrating over the
observed wavelength range.  The line flux for the [O~I] line at 63 $\mu$m
is measured.  Further, the integrated luminosity from the H$_2$ lines
for each SNR is determined.  We estimate the  temperature for the big
grains from fitting of a modified black body. 

\subsection{The MIPS SEDs: [O~I] Line and Continuum Emission } 
\begin{figure}
\includegraphics[width=13.5cm]{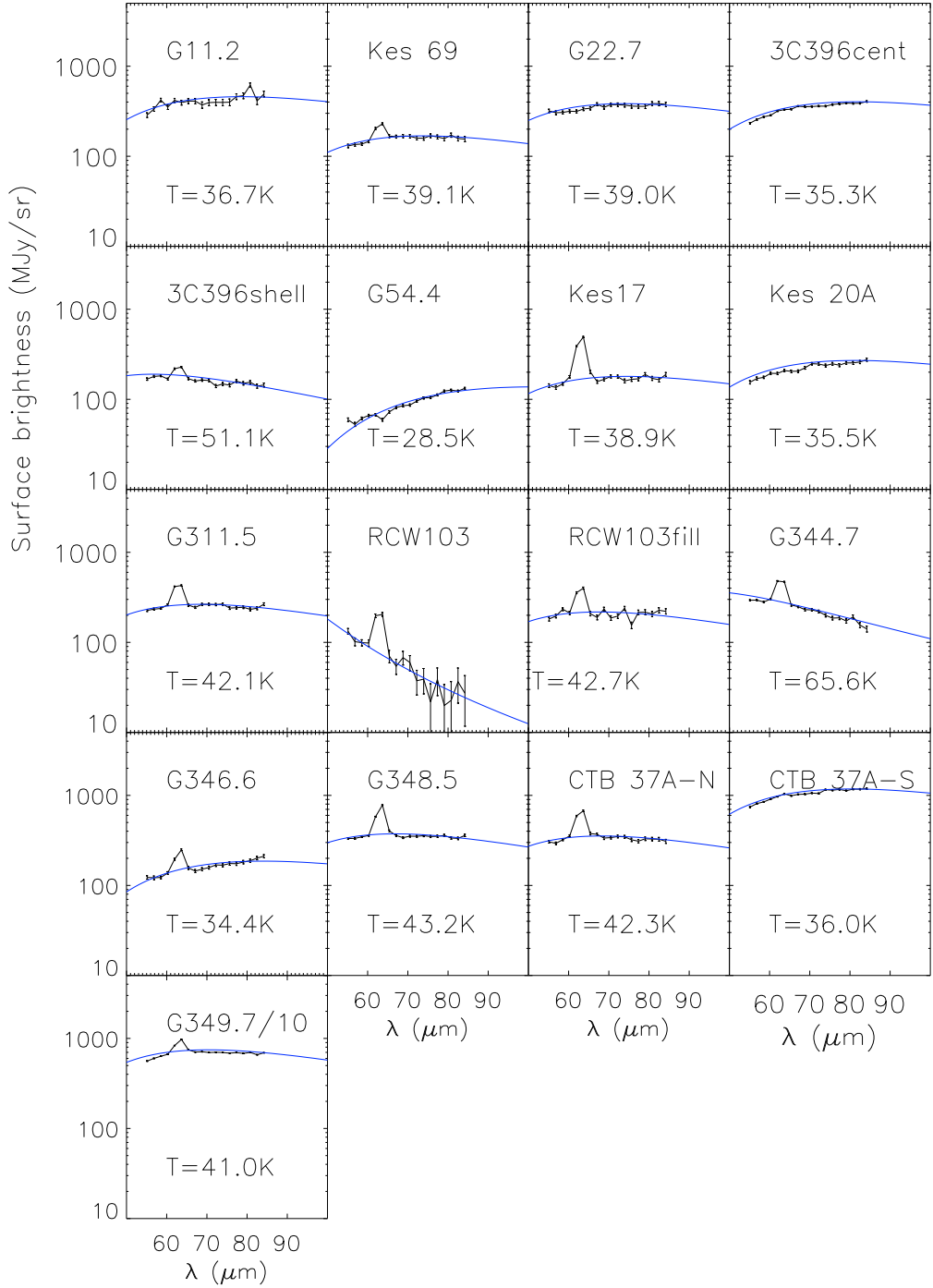}
\caption{The MIPS SEDs for all the SNRs. A modified black body has been fitted to each and the derived temperature is provided in each panel. Note  the temperature of RCW 103 is not given since the background region is contaminated by the emission from the SNR, and 
 the scale of G349+0.2 has been changed by a factor or 10.}
\label{MIPS_SED_all}
\end{figure}

Figure ~\ref{MIPS_SED_all} shows the MIPS SEDs for all the SNRs. Line
emission due to neutral oxygen at 63 $\mu$m is clearly seen in many of
the spectra.  10 SNRs in our sample show clear evidence for [O~I]
emission as shown in Figure~\ref{MIPS_SED_all}.  
A Gaussian fit was performed to the [O~I] lines, and the surface brightnesses of
[O~I] are given in Table~\ref{dust_lum_obs}. 

The [O~I] line luminosity has been calculated 
after extinction correction and accounting for
distances as given in Table~\ref{basic_info}. 
Several SNRs, including RCW 103 show tentative evidence for weak [O~I] emission in the off position. 
If true, the measured [O~I] line would be slightly underestimated, by 10\% or less. 
Typical extinction corrections are less than 10\% at 63 $\mu$m. 

Large  (silicate) grains dominate
the emission at long wavelengths \citep[e.g.][]{desert}.  
The grains are in thermal equilibrium and
we can  estimate their temperature through a fit by a modified black
body, with a dust emissivity index of $\beta=-2$.  The modified black body fits are superposed on the SEDs  in
Figure ~\ref{MIPS_SED_all} and the derived black body temperatures are
provided in Table~\ref{dust_lum_obs}.  We find the range of temperatures
for SNR heated dust to be between 29K and 66K.  
The dust temperature for most of the SNRs is 
between 30 and 43 K. 
The high-temperature exceptions are the shell of 3C396 and G344.7-0.1 with temperatures of 51 and 66 K, respectively. 
Further, for the position RCW103, the background is contaminated by diffuse emission that compromises the SED. 
Although the same background is present for the RCW103 fill position, the emission from the dust is stronger relative to the background.

The temperatures (29-66 K) we derive for the  SNRs are higher  
than a typical ISM temperature of 15--20 K \citep[e.g.][]{reach95,lidraine01}.
The black body temperatures derived show that the dust is heated in
excess of what is expected by the interstellar radiation field.  As
shown in \citet{mathis}, the temperature of the silicate dust scales
roughly as the strength of the interstellar radiation field to the one
sixth power for a dust emissivity of $\beta=-2$.  The derived temperatures thus indicate a heating source
that is 10-800 times stronger than the local interstellar radiation
field.  We discuss in Section 4 the possible origin for the additional
heating.


\subsection{Emission Lines from H$_2$ and Ionic Lines} 

The observed line brightness from H$_2$ and ionic lines are presented in
Table~\ref{lines_obs} and the de--reddened values are presented in
Table~\ref{lines_dered}. The lines were fitted with Gaussian profiles. 
Small wavelength segments were selected and the lines within each segment were fit
simultaneously with the a low order fit to the background. We find H$_2$
emission in 8 of the SNRs, expanding greatly on the sample by
\citet{hewitt} to a total number of 14 SNRs among the sample SNRs in \citep{reach}.
Further, many ionic lines are observed. For G22.7-0.2 there is
tentative evidence for the detection of the H$_2$ S(0) and H$_2$ S(2) lines although they
are very faint relative to the continuum and adjacent PAH features.
 The
total H$_2$ luminosity is estimated in a similar manner. 
The dust--to--H$_2$ luminosity ratio is given in Table \ref{dust_lum_obs}.  


\begin{deluxetable}{cccccccc}
\rotate
\tabletypesize{\scriptsize}
\tablecaption{The dust luminosity, H$_2$ intensity, H$_2$ luminosity, OI intensity, OI luminosity, and the ratio of the H$_2$ luminosity to dust luminosity. The dust luminosity is integrated from 5 to 80 $\mu$m and is interpolated between 35 and 55 $\mu$m. The errors on the luminosity are of the order 10\% and are mainly due to systematic 
uncertainties in the background subtraction. The uncertainty on the black--body temperatures are 1K. }
\tablehead{\colhead{SNR} &  \colhead{Dust luminosity} & \colhead{H$_2$ brightness} & \colhead{H$_2$ luminosity} & \colhead{OI brightness} & \colhead{OI luminosity} & \colhead{$\frac{L(H_2)}{L(dust)}$} & \colhead{$\mathrm{T_{BB}}$}\\
\colhead{} & \colhead{L$_\odot$} & \colhead{$\mathrm{erg/s/cm^2/sr}$} & \colhead{L$_\odot$} & \colhead{$\mathrm{erg/s/cm^2/sr}$} & \colhead{L$_\odot$} & & \colhead{K}}
\startdata
G11.2&         291&2.2(0.1)E-3&3.9(0.2)E 0&0.0(0.0)E 0&0.0(0.0)E 0&1.3(0.1)E-2&
          38\\
Kes 69&         118&1.5(0.2)E-3&2.4(0.2)E 0&1.9(0.1)E-4&2.2(0.1)E 0&2.1(0.2)E-2&
          39\\
G22.7&         211&7.6(2.1)E-5&2.1(0.5)E-2&0.0(0.0)E 0&0.0(0.0)E 0&10(2.4)E-5&
          40\\
3C396cent&         473&1.9(0.1)E-5&1.6(0.1)E-2&0.0(0.0)E 0&0.0(0.0)E 0&
3.4(0.4)E-5&          35\\
3C396shell&         301&1.6(0.2)E-3&3.2(0.2)E 0&1.5(4.9)E-4&2.8(9.2)E 0&
1.1(0.1)E-2&          51\\
G54.4&          22&3.6(1.3)E-5&6.5(1.7)E-3&0.0(0.0)E 0&0.0(0.0)E 0&3.0(0.8)E-4&
          29\\
Kes 17&         460&4.8(0.3)E-3&2.7(0.1)E 1&8.1(0.6)E-4&3.3(0.2)E 1&5.8(0.6)E-2&
          39\\
Kes 20A&        2034&3.9(0.8)E-5&1.5(0.2)E-1&0.0(0.0)E 0&0.0(0.0)E 0&7.4(1.3)E-5
&          35\\
G311.5&         948&1.6(0.2)E-3&2.2(0.2)E 1&4.9(1.1)E-4&4.6(1.0)E 1&2.3(0.3)E-2&
          42\\
RCW103&          53&2.6(0.3)E-3&1.5(0.1)E 0&3.2(5.0)E-4&1.4(2.3)E 0&2.8(0.4)E-2&
         ...\\
RCW103fill&          60&1.9(1.0)E-4&1.2(1.0)E-1&4.7(5.9)E-4&2.1(2.7)E 0&
1.9(1.7)E-3&          43\\
G344.7&        1633&1.2(0.1)E-3&1.3(0.1)E 1&6.0(1.0)E-4&5.2(0.9)E 1&7.8(1.0)E-3&
          66\\
G346.6&         434&1.2(0.1)E-3&9.0(0.5)E 0&2.4(0.3)E-4&1.3(0.2)E 1&2.1(0.2)E-2&
          34\\
G348.5&        1902&8.9(1.8)E-4&1.2(0.2)E 1&9.4(0.6)E-4&7.5(0.5)E 1&6.1(1.0)E-3&
          43\\
CTB 37A-N&        1122&2.1(0.2)E-3&1.6(0.1)E 1&8.3(1.7)E-4&4.3(0.9)E 1&
1.4(0.2)E-2&          42\\
CTB 37A-S&        3604&3.8(1.6)E-5&9.2(3.1)E-2&0.0(0.0)E 0&0.0(0.0)E 0&
2.5(0.9)E-5&          36\\
G349.7&       77998&1.7(0.5)E-3&4.3(1.4)E 1&7.0(0.3)E-3&1.5(0.1)E 3&5.6(1.8)E-4&
          41\\
\enddata
\label{dust_lum_obs}
\end{deluxetable}

\begin{deluxetable}{ccccccccccccccccc}
\tabletypesize{\tiny}
\tablecaption{The observed line brightness ($erg~s^{-1}~cm^{-2}~sr^{-1}$) for molecular hydrogen and atomic lines. Local background has been subtracted from the spectra.}
\tablehead{\colhead{Transition}&\colhead{$\lambda$ ($\mu$m)}&\colhead{G11.2}&
\colhead{G22.7}&\colhead{3C396cent}&\colhead{Kes 20A}&\colhead{RCW103}&
\colhead{RCW103fill}}
\startdata
  H$_2$ S(0) &       28.22 & 0.0(0.0)E 0 & 0.0(0.0)E 0 & 0.0(0.0)E 0 & 5.9(2.5)E-6 & 0.0(0.0)E 0 & 0.0(0.0)E 0\\
  H$_2$ S(1) &       17.04 & 1.7(0.4)E-5 & 1.9(0.3)E-5 & 1.2(0.1)E-5 & 1.2(0.1)E-5 & 9.3(0.7)E-5 & 3.8(0.4)E-5\\
  H$_2$ S(2) &       12.28 & 4.3(0.6)E-5 & 0.0(0.0)E 0 & 0.0(0.0)E 0 & 0.0(0.0)E 0 & 9.5(1.0)E-5 & 0.0(0.0)E 0\\
  H$_2$ S(3) &        9.67 & 1.5(0.1)E-4 & 1.4(0.5)E-5 & 0.0(0.0)E 0 & 0.0(0.0)E 0 & 4.5(0.7)E-4 & 0.0(0.0)E 0\\
  H$_2$ S(4) &        8.03 & 2.0(0.2)E-4 & 0.0(0.0)E 0 & 0.0(0.0)E 0 & 0.0(0.0)E 0 & 2.4(0.4)E-4 & 0.0(0.0)E 0 &          \\
  H$_2$ S(5) &        6.91 & 6.6(0.3)E-4 & 0.0(0.0)E 0 & 0.0(0.0)E 0 & 0.0(0.0)E 0 & 8.1(0.6)E-4 & 8.8(3.0)E-5 &          \\
  H$_2$ S(6) &        6.11 & 1.8(0.3)E-4 & 0.0(0.0)E 0 & 0.0(0.0)E 0 & 0.0(0.0)E 0 & 1.6(0.6)E-4 & 0.0(0.0)E 0 &          \\
  H$_2$ S(7) &        5.51 & 5.6(0.2)E-4 & 0.0(0.0)E 0 & 0.0(0.0)E 0 & 0.0(0.0)E 0 & 4.9(0.5)E-4 & 5.2(5.8)E-5 &          \\
 $[$FeII$]$  &        5.34 & 0.0(0.0)E 0 & 0.0(0.0)E 0 & 0.0(0.0)E 0 & 7.6(8.2)E-6 & 1.0(0.1)E-3 & 1.3(0.0)E-3\\
 $[$ArII$]$  &        6.98 & 6.1(2.7)E-5 & 0.0(0.0)E 0 & 2.2(2.4)E-5 & 0.0(0.0)E 0 & 7.8(0.7)E-4 & 5.5(0.5)E-4\\
$[$ArIII$]$  &        9.00 & 0.0(0.0)E 0 & 0.0(0.0)E 0 & 0.0(0.0)E 0 & 0.0(0.0)E 0 & 0.0(0.0)E 0 & 0.0(0.0)E 0\\
 $[$NeII$]$  &        12.8 & 6.2(0.6)E-5 & 4.9(1.5)E-5 & 9.4(1.1)E-5 & 0.0(0.0)E 0 & 1.2(0.0)E-3 & 9.4(0.2)E-4 &          \\
$[$NeIII$]$  &        15.5 & 1.8(0.1)E-4 & 0.0(0.0)E 0 & 0.0(0.0)E 0 & 0.0(0.0)E 0 & 7.8(0.2)E-4 & 3.5(0.1)E-4\\
 $[$FeII$]$  &        17.9 & 6.2(0.5)E-5 & 0.0(0.0)E 0 & 0.0(0.0)E 0 & 0.0(0.0)E 0 & 3.0(0.1)E-4 & 2.1(0.1)E-4\\
 $[$SIII$]$  &        18.7 & 6.0(0.5)E-5 & 1.1(0.3)E-5 & 1.7(0.2)E-5 & 2.5(0.8)E-6 & 2.5(0.1)E-4 & 8.9(0.4)E-5 &          \\
 $[$FeII$]$  &        24.5 & 3.9(0.8)E-5 & 0.0(0.0)E 0 & 0.0(0.0)E 0 & 0.0(0.0)E 0 & 7.9(1.3)E-5 & 3.7(0.6)E-5 &          \\
   $[$SI$]$  &        25.2 & 1.9(0.6)E-5 & 1.2(0.9)E-6 & 4.3(0.4)E-5 & 1.9(0.6)E-6 & 2.2(1.2)E-5 & 1.4(0.3)E-5\\
 $[$FeII$]$  &        25.9 & 2.7(0.1)E-4 & 6.2(1.8)E-6 & 4.2(1.5)E-6 & 2.4(1.0)E-6 & 5.6(0.2)E-4 & 3.2(0.1)E-4 &          \\
 $[$SIII$]$  &        33.5 & 3.6(0.4)E-5 & 0.0(0.0)E 0 & 1.0(0.0)E-4 & 0.0(0.0)E 0 & 1.6(0.1)E-4 & 5.2(0.7)E-5\\
     $[$SiII$]$  &        34.8 & 5.5(0.5)E-5 & 3.5(0.6)E-5 & 4.3(0.3)E-5 & 1.0(0.3)E-5 & 6.5(0.2)E-4 & 4.8(0.1)E-4\\
 $[$FeII$]$  &       35.35 & 1.8(0.3)E-5 & 0.0(0.0)E 0 & 0.0(0.0)E 0 & 0.0(0.0)E 0 & 1.1(0.1)E-4 & 7.1(0.8)E-5 &          \\
   $[$OI$]$  &          63 & 0.0(0.0)E 0 & 0.0(0.0)E 0 & 0.0(0.0)E 0 & 0.0(0.0)E 0 & 3.2(5.0)E-4 & 4.7(5.9)E-4 &          \\
\hline
& &G311.5&G344.7&CTB 37A-N&CTB 37A-S&G54.4\\
  H$_2$ S(0) &       28.22 & 0.0(0.0)E 0 & 2.0(0.8)E-6 & 9.2(1.2)E-6 & 1.5(0.3)E-5 & 6.7(0.4)E-6\\
  H$_2$ S(1) &       17.04 & 6.0(0.2)E-5 & 2.1(0.2)E-5 & 3.0(0.2)E-5 & 1.3(0.8)E-5 & 1.1(0.6)E-5\\
  H$_2$ S(2) &       12.28 & 5.5(0.4)E-5 & 3.2(0.5)E-5 & 6.4(1.8)E-5 & 0.0(0.0)E 0 & 1.5(0.6)E-5\\
  H$_2$ S(3) &        9.67 & 8.1(0.9)E-5 & 4.6(0.6)E-5 & 1.2(0.0)E-4 & 0.0(0.0)E 0 & 0.0(0.0)E 0\\
  H$_2$ S(4) &        8.03 & 1.4(0.1)E-4 & 6.3(1.1)E-5 & 1.7(0.3)E-4 & 0.0(0.0)E 0 & 0.0(0.0)E 0\\
  H$_2$ S(5) &        6.91 & 5.0(0.4)E-4 & 2.5(0.2)E-4 & 5.6(0.3)E-4 & 0.0(0.0)E 0 & 0.0(0.0)E 0\\
  H$_2$ S(6) &        6.11 & 1.2(0.4)E-4 & 5.9(1.3)E-5 & 1.6(0.4)E-4 & 0.0(0.0)E 0 & 0.0(0.0)E 0\\
  H$_2$ S(7) &        5.51 & 3.2(0.5)E-4 & 1.5(0.2)E-4 & 3.8(0.5)E-4 & 0.0(0.0)E 0 & 0.0(0.0)E 0\\
 $[$FeII$]$  &        5.34 & 5.0(0.5)E-4 & 5.8(0.2)E-4 & 5.1(0.6)E-4 & 0.0(0.0)E 0 & 1.3(1.3)E-5\\
 $[$ArII$]$  &        6.98 & 1.2(0.4)E-4 & 3.3(0.2)E-4 & 1.9(0.2)E-4 & 4.7(2.4)E-5 & 0.0(0.0)E 0\\
$[$ArIII$]$  &        9.00 & 0.0(0.0)E 0 & 1.9(0.5)E-5 & 0.0(0.0)E 0 & 0.0(0.0)E 0 & 0.0(0.0)E 0\\
   pNeII$]$  &        12.8 & 1.5(0.1)E-4 & 4.8(0.1)E-4 & 5.2(0.2)E-4 & 1.8(0.2)E-4 & 2.2(0.7)E-5\\
$[$NeIII$]$  &        15.5 & 4.4(0.2)E-5 & 2.9(0.0)E-4 & 1.7(0.0)E-4 & 1.4(0.7)E-5 & 3.3(1.5)E-6\\
 $[$FeII$]$  &        17.9 & 2.0(0.1)E-5 & 1.2(0.0)E-4 & 0.0(0.0)E 0 & 0.0(0.0)E 0 & 0.0(0.0)E 0\\
  $[$SIII$]$ &        18.7 & 1.7(0.1)E-5 & 6.1(0.2)E-5 & 0.0(0.0)E 0 & 8.2(0.9)E-5 & 2.4(1.0)E-6\\
 $[$FeII$]$  &        24.5 & 2.8(1.5)E-6 & 3.2(0.4)E-5 & 6.5(6.3)E-6 & 0.0(0.0)E 0 & 0.0(0.0)E 0\\
   $[$SI$]$  &        25.2 & 4.4(0.7)E-6 & 8.2(2.7)E-6 & 1.6(0.2)E-5 & 4.9(4.8)E-6 & 0.0(0.0)E 0\\
 $[$FeII$]$  &        25.9 & 5.6(0.2)E-5 & 3.0(0.1)E-4 & 1.6(0.1)E-4 & 1.9(0.5)E-5 & 1.2(0.5)E-6\\
 $[$SIII$]$  &        33.5 & 1.6(0.3)E-5 & 5.1(0.5)E-5 & 3.0(0.7)E-5 & 7.6(0.7)E-5 & 1.3(0.1)E-5\\
 $[$SiII$]$  &        34.8 & 3.0(0.1)E-4 & 4.1(0.1)E-4 & 9.4(0.2)E-4 & 2.8(0.1)E-4 & 2.0(0.1)E-5\\
 $[$FeII$]$  &       35.35 & 1.8(0.4)E-5 & 6.1(1.7)E-5 & 6.8(0.9)E-5 & 0.0(0.0)E 0 & 0.0(0.0)E 0\\
   $[$OI$]$  &          63 & 4.9(1.1)E-4 & 6.0(1.0)E-4 & 8.3(1.7)E-4 & 0.0(0.0)E 0 & 0.0(0.0)E 0\\
\enddata
\label{lines_obs}
\end{deluxetable}
\begin{deluxetable}{cccccccccc}
\tabletypesize{\tiny}
\tablecaption{The de-reddened line brightness ($erg~s^{-1}~cm^{-2}~sr^{-1}$) for molecular hydrogen and atomic lines. Local background has been subtracted from the spectra and an extinction law with R$_\mathrm{V}=3.1$.}
\tablehead{\colhead{Transition}&\colhead{$\lambda$ ($\mu$m)}&\colhead{G11.2}&
\colhead{G22.7}&\colhead{3C396cent}&\colhead{Kes 20A}&\colhead{RCW103}&
\colhead{RCW103fill}}

\startdata
  H$_2$ S(0) &       28.22 & 0.0(0.0)E 0 & 0.0(0.0)E 0 & 0.0(0.0)E 0 & 1.3(0.6)E-5 & 0.0(0.0)E 0 & 0.0(0.0)E 0\\
  H$_2$ S(1) &       17.04 & 2.1(0.5)E-5 & 2.7(0.5)E-5 & 1.7(0.1)E-5 & 2.5(0.3)E-5 & 9.9(0.8)E-5 & 4.1(0.4)E-5\\
  H$_2$ S(2) &       12.28 & 5.0(0.7)E-5 & 0.0(0.0)E 0 & 0.0(0.0)E 0 & 0.0(0.0)E 0 & 9.9(1.0)E-5 & 0.0(0.0)E 0\\
  H$_2$ S(3) &        9.67 & 2.0(0.1)E-4 & 2.3(0.8)E-5 & 0.0(0.0)E 0 & 0.0(0.0)E 0 & 4.9(0.7)E-4 & 0.0(0.0)E 0\\
  H$_2$ S(4) &        8.03 & 4.1(0.3)E-4 & 0.0(0.0)E 0 & 0.0(0.0)E 0 & 0.0(0.0)E 0 & 3.1(0.5)E-4 & 0.0(0.0)E 0 &          \\
  H$_2$ S(5) &        6.91 & 8.8(0.4)E-4 & 0.0(0.0)E 0 & 0.0(0.0)E 0 & 0.0(0.0)E 0 & 9.0(0.7)E-4 & 9.8(3.4)E-5 &          \\
  H$_2$ S(6) &        6.11 & 2.3(0.4)E-4 & 0.0(0.0)E 0 & 0.0(0.0)E 0 & 0.0(0.0)E 0 & 1.8(0.6)E-4 & 0.0(0.0)E 0 &          \\
  H$_2$ S(7) &        5.51 & 6.5(0.3)E-4 & 0.0(0.0)E 0 & 0.0(0.0)E 0 & 0.0(0.0)E 0 & 5.1(0.5)E-4 & 5.5(6.2)E-5 &          \\
 $[$FeII$]$  &        5.34 & 0.0(0.0)E-5 & 0.0(0.0)E-5 & 0.0(0.0)E-4 & 1.7(1.8)E-5 & 1.1(0.1)E-3 & 1.4(0.0)E-3\\
 $[$ArII$]$  &        6.98 & 7.1(3.1)E-5 & 0.0(0.0)E 0 & 3.1(3.4)E-5 & 0.0(0.0)E 0 & 8.2(0.7)E-4 & 5.8(0.5)E-4\\
$[$ArIII$]$  &        9.00 & 0.0(0.0)E 0 & 0.0(0.0)E 0 & 0.0(0.0)E 0 & 0.0(0.0)E 0 & 0.0(0.0)E 0 & 0.0(0.0)E 0\\
 $[$NeII$]$  &        12.8 & 7.9(0.8)E-5 & 7.9(2.4)E-5 & 1.7(0.2)E-4 & 0.0(0.0)E 0 & 1.3(0.0)E-3 & 1.0(0.0)E-3 &          \\
$[$NeIII$]$  &        15.5 & 2.2(0.1)E-4 & 0.0(0.0)E 0 & 0.0(0.0)E 0 & 0.0(0.0)E 0 & 8.3(0.2)E-4 & 3.7(0.1)E-4\\
 $[$FeII$]$  &        17.9 & 8.2(0.7)E-5 & 0.0(0.0)E-5 & 0.0(0.0)E-4 & 0.0(0.0)E-4 & 3.2(0.1)E-4 & 2.3(0.1)E-4\\
 $[$SIII$]$  &        18.7 & 7.9(0.6)E-5 & 1.8(0.4)E-5 & 3.1(0.3)E-5 & 7.8(2.5)E-6 & 2.7(0.1)E-4 & 9.8(0.4)E-5 &          \\
 $[$FeII$]$  &        24.5 & 4.6(1.0)E-5 & 0.0(0.0)E-5 & 0.0(0.0)E-5 & 0.0(0.0)E-5 & 8.3(1.4)E-5 & 4.0(0.6)E-5 &          \\
   $[$SI$]$  &        25.2 & 2.3(0.7)E-5 & 1.7(1.3)E-6 & 6.4(0.6)E-5 & 3.8(1.2)E-6 & 2.3(1.3)E-5 & 1.5(0.3)E-5\\
 $[$FeII$]$  &        25.9 & 3.2(0.1)E-4 & 8.4(2.4)E-6 & 6.1(2.2)E-6 & 4.7(2.0)E-6 & 5.9(0.2)E-4 & 3.3(0.1)E-4 &          \\
 $[$SIII$]$  &        33.5 & 4.1(0.5)E-5 & 0.0(0.0)E 0 & 1.3(0.1)E-4 & 0.0(0.0)E 0 & 1.7(0.2)E-4 & 5.5(0.7)E-5\\
 $[$SiII$]$  &        34.8 & 6.1(0.6)E-5 & 4.3(0.7)E-5 & 5.6(0.3)E-5 & 1.6(0.5)E-5 & 6.8(0.2)E-4 & 5.0(0.1)E-4\\
 $[$FeII$]$  &       35.35 & 2.0(0.3)E-5 & 0.0(0.0)E-5 & 0.0(0.0)E-5 & 0.0(0.0)E-5 & 1.2(0.2)E-4 & 7.4(0.8)E-5 &          \\
   $[$OI$]$  &          63 & 0.0(0.0)E 0 & 0.0(0.0)E 0 & 0.0(0.0)E-4 & 0.0(0.0)E-4 & 3.2(5.0)E-4 & 4.8(6.0)E-4 &          \\
\hline
& &G311.5&G344.7&CTB 37A-N&CTB 37A-S&G54.4\\
  H$_2$ S(0) &       28.22 & 0.0(0.0)E 0 & 3.2(1.4)E-6 & 1.2(0.2)E-5 & 2.1(0.5)E-5 & 7.4(0.5)E-6\\
  H$_2$ S(1) &       17.04 & 7.3(0.2)E-5 & 3.4(0.3)E-5 & 4.0(0.3)E-5 & 1.8(1.1)E-5 & 1.2(0.7)E-5\\
  H$_2$ S(2) &       12.28 & 6.4(0.5)E-5 & 4.6(0.7)E-5 & 8.0(2.2)E-5 & 0.0(0.0)E 0 & 1.6(0.6)E-5\\
  H$_2$ S(3) &        9.67 & 1.1(0.1)E-4 & 9.5(1.2)E-5 & 1.8(0.1)E-4 & 0.0(0.0)E 0 & 0.0(0.0)E 0\\
  H$_2$ S(4) &        8.03 & 3.2(0.2)E-4 & 4.7(0.8)E-4 & 5.3(0.8)E-4 & 0.0(0.0)E 0 & 0.0(0.0)E 0\\
  H$_2$ S(5) &        6.91 & 6.9(0.6)E-4 & 5.5(0.4)E-4 & 8.9(0.5)E-4 & 0.0(0.0)E 0 & 0.0(0.0)E 0\\
  H$_2$ S(6) &        6.11 & 1.6(0.5)E-4 & 1.2(0.3)E-4 & 2.4(0.5)E-4 & 0.0(0.0)E 0 & 0.0(0.0)E 0\\
  H$_2$ S(7) &        5.51 & 3.8(0.6)E-4 & 2.3(0.2)E-4 & 4.8(0.6)E-4 & 0.0(0.0)E 0 & 0.0(0.0)E 0\\
 $[$FeII$]$  &        5.34 & 6.2(0.6)E-4 & 9.9(0.4)E-4 & 7.0(0.8)E-4 & 0.0(0.0)E-4 & 1.4(1.4)E-5\\
 $[$ArII$]$  &        6.98 & 1.4(0.4)E-4 & 4.9(0.3)E-4 & 2.3(0.2)E-4 & 5.9(3.0)E-5 & 0.0(0.0)E 0\\
$[$ArIII$]$  &        9.00 & 0.0(0.0)E 0 & 9.1(2.4)E-5 & 0.0(0.0)E 0 & 0.0(0.0)E 0 & 0.0(0.0)E 0\\
 $[$NeII$]$  &        12.8 & 2.0(0.1)E-4 & 9.3(0.2)E-4 & 7.7(0.3)E-4 & 2.6(0.3)E-4 & 2.5(0.7)E-5\\
$[$NeIII$]$  &        15.5 & 5.5(0.2)E-5 & 5.1(0.1)E-4 & 2.3(0.0)E-4 & 2.0(0.9)E-5 & 3.6(1.6)E-6\\
 $[$FeII$]$  &        17.9 & 2.6(0.1)E-5 & 2.5(0.1)E-4 & 0.0(0.0)E-5 & 0.0(0.0)E-5 & 0.0(0.0)E-4\\
 $[$SIII$]$  &        18.7 & 2.2(0.1)E-5 & 1.3(0.0)E-4 & 0.0(0.0)E 0 & 1.3(0.1)E-4 & 2.7(1.2)E-6\\
 $[$FeII$]$  &        24.5 & 3.4(1.8)E-6 & 5.1(0.6)E-5 & 8.5(8.2)E-6 & 0.0(0.0)E-6 & 0.0(0.0)E-4\\
   $[$SI$]$  &        25.2 & 5.3(0.8)E-6 & 1.3(0.4)E-5 & 2.1(0.2)E-5 & 6.4(6.2)E-6 & 0.0(0.0)E 0\\
 $[$FeII$]$  &        25.9 & 6.6(0.3)E-5 & 4.7(0.1)E-4 & 2.0(0.1)E-4 & 2.4(0.6)E-5 & 1.3(0.6)E-6\\
 $[$SIII$]$  &        33.5 & 1.8(0.3)E-5 & 7.1(0.7)E-5 & 3.6(0.9)E-5 & 9.2(0.9)E-5 & 1.4(0.1)E-5\\
 $[$SiII$]$  &        34.8 & 3.3(0.1)E-4 & 5.5(0.1)E-4 & 1.1(0.0)E-3 & 3.3(0.1)E-4 & 2.1(0.1)E-5\\
 $[$FeII$]$  &       35.35 & 2.0(0.5)E-5 & 8.2(2.3)E-5 & 8.1(1.1)E-5 & 0.0(0.0)E-5 & 0.0(0.0)E-4\\
   $[$OI$]$  &          63 & 5.1(1.1)E-4 & 6.4(1.1)E-4 & 8.6(1.8)E-4 & 0.0(0.0)E-4 & 0.0(0.0)E-3\\
\enddata
\label{lines_dered}
\end{deluxetable}




\subsection{Integrated Dust Luminosity}
\begin{figure}
\includegraphics[width=15cm]{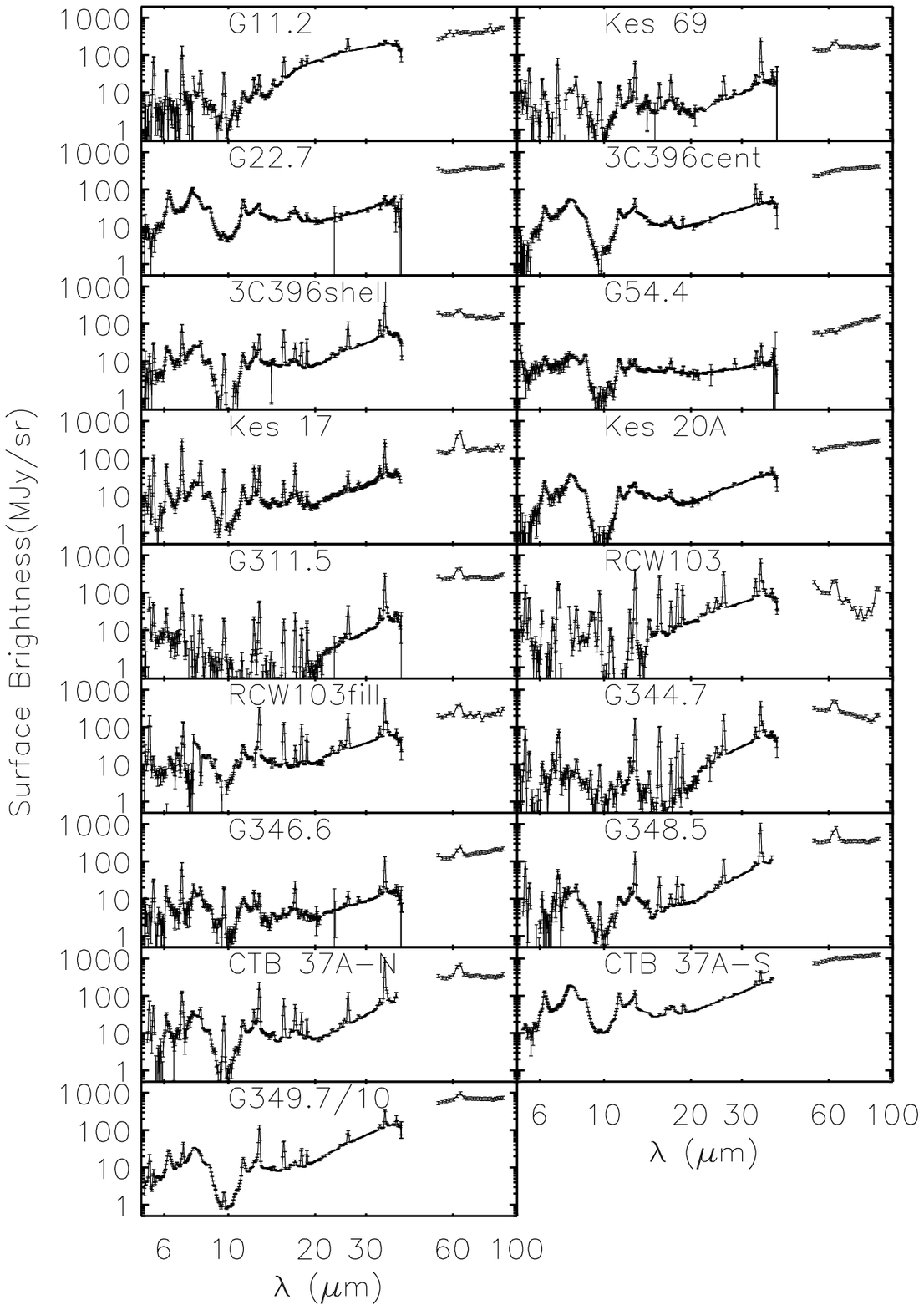}
\caption{The IRS spectra and MIPS SEDs for the SNRs before line subtraction and de--reddening. 
Note the scale of G349.7+0.2 has been changed by a factor of 10.}
\label{just_spectra}
\end{figure}

Figure~\ref{just_spectra} shows the {\it Spitzer} IRS spectra and MIPS SED
observations of the SNRs before dereddening.   The spectra show  a rise of
the continuum longwards of 20 $\mu$m with a peak typically in the
70-85$\mu$m range. There are several indications for PAH emission in the
short wavelength part of the spectrum, including the features at 11.3
$\mu$m, 16--19 $\mu$m as well as 6.2, 7.7, and 8.6$\mu$m. The features are
most noticeable in e.g. Kes 17, CTB 37A, and RCW 103. It is thus further
likely that the mid--infrared continuum emission is due to PAH emission, as
predicted by e.g. the DUSTEM models. 


We have integrated the dust emission from the observed part of the SED. 
All the spectra have been de-reddened using the foreground hydrogen column
given in Table~\ref{basic_info} and using the extinction data provided
by \citet{draine}, assuming $R_V=3.1$.  
The luminosity is calculated
based on a solid angle of the MIPS SED (2$\times$3 pixels, or
19.6\arcsec$\times$30\arcsec) and thus assumes the emission is uniform over this scale. 
 The luminosities are given in
Table~\ref{dust_lum_obs}.   

The ratio between the H$_2$ and dust luminosities for each SNR is  given in
Table~\ref{dust_lum_obs}.
For the clear detection cases of H$_2$ lines
(G11.2-0.3, Kes 69, 3C396 shell, Kes 17, G311.5-0.3, G344.7-0.1,
G346.6-0.2, G348.5-0.0,
CTB 37A--N) the ratio ranges from 0.6\% to 6\%. 
These ratios are  not as high as 17\% observed in radio galaxies \citep{ogle}
or 30\% in the IGM shock in Stephan's Quintet \citep{appleton}, but
still higher than those from normal galaxies \citep{soifer}. 
Thus, the main coolant in the interacting SNRs is in the form of infrared 
continuum cooling from the dust.

The background estimate for RCW 103 is very uncertain and has been excluded from 
the sample. 
G349.7+0.2 shows a very small ratio of 6.09$\times10^{-4}$. 
At long wavelength there is a slight contamination from the central object that is also very bright at 
70 $\mu$m. 
The contamination from the point source has been estimated in the 70$\mu$m images through 
point--spread--function fitting to the central source and subtracting it. The contamination is less than 10\% 
and thus not the origin of the low ratio. 


\subsection{ [O I] 63\micron\ Cooling Line}

The 63 $\mu$m line of atomic oxygen has been observed in a few
interacting SNRs previously. In IC443, a good spatial correlation with
H$_2$ 1-0 S(1) line emission was observed \citep{burton90}. This
correlation between H$_2$ and [O I] line emission suggests an origin
from the same shock. Atomic oxygen emission was also detected toward W44
and 3C391 between (0.3--1.4)$\times$10$^{-3}$ \ergflux\ sr$^{-1}$
\citep{reachrho96}. The [O I] lines appear brightest at the edge of the
remnant which also favors a shock origin. IRS observations of the SNRs
presented here are also located at the shock front where IR emission
peaks and are likely located at or near the peak in emission from atomic
oxygen. We typically find line brightnesses of (1--10)$\times$10$^{-4}$
\ergflux\ sr$^{-1}$ for the [O I] 63\micron\ line, with the exception of
G349.7+0.2 which is exceptionally bright at 7$\times$10$^{-3}$
\ergflux\ sr$^{-1}$.

Given the correlation of [O~I] with H$_2$ emission observed for other
remnants, a C-shock origin for the atomic oxygen emission seems
plausible. A large parameter space was explored for C-shocks to try to
reproduce the observed [O I] line in IC443 \citep{burton90}. Fast
C-shocks into moderately dense gas which explain the hot H$_2$ gas can
produce significant [O I] emission, but only if oxygen chemistry is
suppressed preventing oxygen from being converted into water or
hydroxyl. This is not thought to be the case here as significant columns of
post-shock OH are observed for many of these remnants \citep[e.g.][]{frail}. Alternatively a
slow, dense C-shock (10\kms , 10$^5$ cm$^{-3}$) produces intense atomic
oxygen emission and fits the derived parameters of the warm H$_2$ gas.

\section{Analysis of Dust Continuum} 

\subsection{Dust Heating}
The dust can  mainly  be heated through two mechanisms: radiative and
collisional heating.  The dominant heating source depends on the dust
particles, and the physical conditions, in particular the temperature
and electron density.  Heating through radiation is given by
\citep[e.g.][]{dwekreview} 
\begin{equation}
H_{rad}=\int F(\lambda) ~Q_{abs}(\lambda) ~ d\lambda
\end{equation}
where $F(\lambda)$ is the radiation field and Q$_{abs}$ is the wavelength dependent
absorption coefficient of the grains.  To first order, Q$_{abs}$ is for a
spherical grain described by $\pi a^28a/\lambda\propto a^3/\lambda$ where
$a$  is the grain size. 
More detailed absorption coefficients, which takes the detailed properties of carbon and silicate grains and into account,  are given in e.g. \citet{draine}. 
The relationship is then more complicated than a simple power-law dependence on grain size that nevertheless serves as a useful reference for the general behaviour of the radiation field. 

The radiative heating comes from at least two terms for the molecular interacting
SNRs: one is the diffuse stellar radiation, similar to what is observed
in the solar neighborhood and the other is hydrogen recombination.  The
IRS spectra indicate that there is a gas component (for example, Fe-emitting gas) next to or associated with
the dust that is relatively warm, 5000 -10\,000 K and which has an electron
density, $n_e$, of 100--1000 $cm^{-3}$ \citep{hewitt}.  The energy emitted from
the recombination will heat the dust and the effect can be estimated based on the recombination rate. 

We assume the electron density is the same as the ionized
hydrogen density, that is, all the electrons are from hydrogen.  
 This is a good
approximation at the relatively low  temperature as observed here since most metals are only singly or weakly ionized.  
The radiation field from recombination is then given by 
\begin{equation}
F(\lambda)=E_{rec} n_e n_0 \alpha V/\Omega, 
\end{equation}
where $E_{rec}=13.6$eV, $V$ is the gas volume and $\Omega$ is the area of the shock that
is covered by the dust, and $n_0$ is the density of neutral hydrogen. 
We adopt the recombination coefficient from
\citet{osterbrock} for case B and for a typical temperature of 10\, 000 K,
$\alpha=2\cdot 10^{-13} cm^3~s^{-1}$.  
The geometric term $V/\Omega$ is uncertain and will depend in detail on
each SNR and the structure of the molecular cloud. 
Here, we for simplicity assume the ionized gas and the dust cover  the same area and the
geometric factor is then reduced to the width, $l$,  of the shock front, which was
estimated in \citet{shull} to be of the order  $10^{16}cm$. The pre--shock
densities determined for the SNRs discussed here can be  $\sim10^4~cm^{-3}$ \citep{hewitt}.
Thus the  incident flux on a dust particle is 
$F(\lambda)=0.2\times(l/10^{ 16}cm)\times(n_e/500cm^{-3})\times(n_0/10^4cm^{-3})~erg~s^{-1}~cm^{-2}$. 
The corresponding energy from the interstellar radiation field is $0.0217
~erg s^{-1} cm^{-2}$ in the solar neighborhood. Recombination can thus be a major
energy source for the dust heating and a  substantial enhancement over the
local interstellar radiation field. Note that the densities and
temperatures derived in \citet{hewitt} are relatively uncertain.  The
temperature is not well constrained without higher energy transitions from,
for example, the 1.64 $\mu$m iron line (its upper level energy is 11,400
K). Further, due to the relatively high dependence on the electron density,
we would expect a larger contribution to the heating from recombination.
Near--infrared observations tracing higher temperature lines will help in
further constraining the amount of heating from recombination. Due to the
uncertainties in calculating the total heating from recombination, we have
decided to leave the strength of the radiation field a free parameter in
the fitting.

\begin{figure}
\includegraphics[width=10cm]{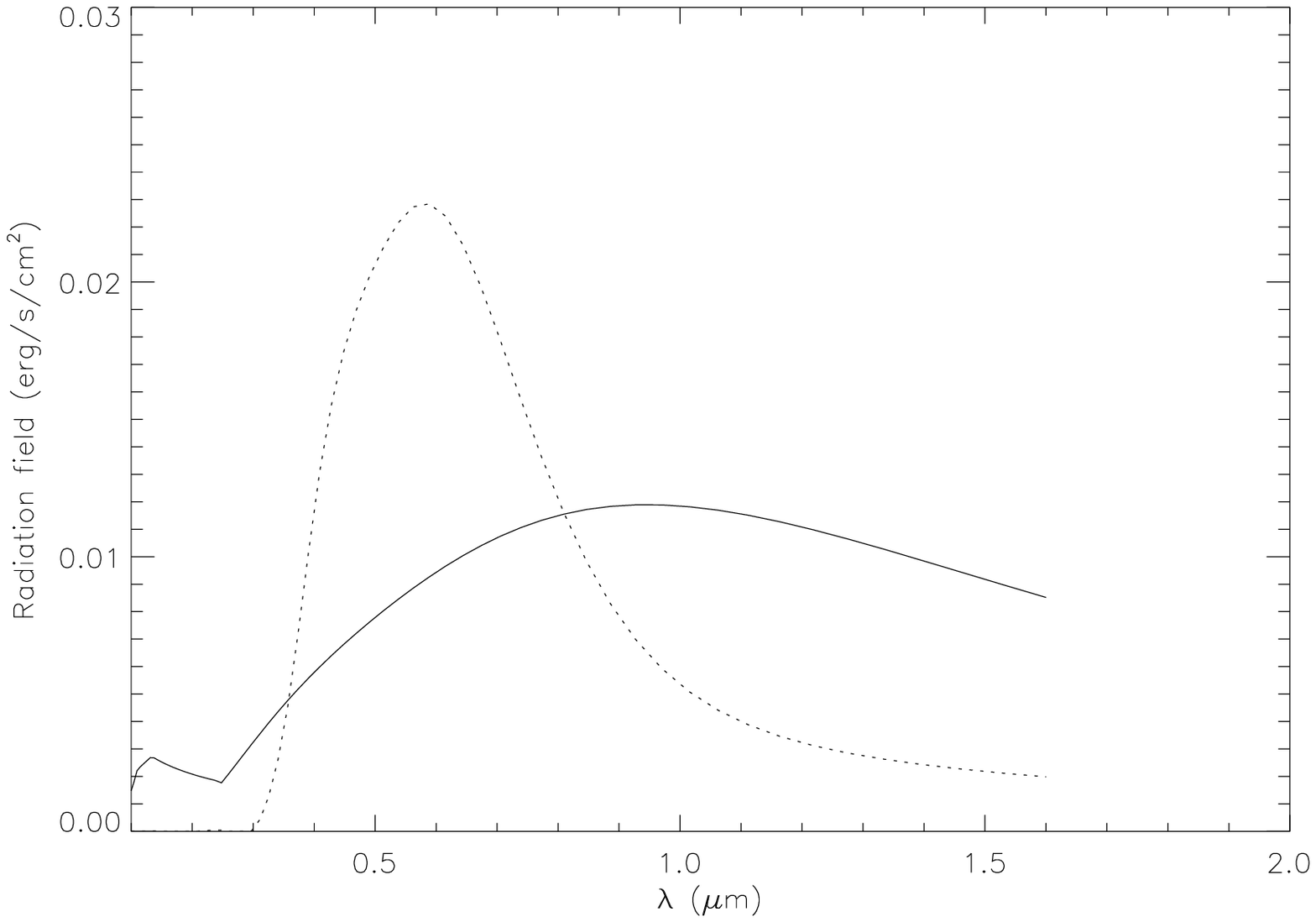}
\caption{Two types of radiation field are shown; the solid
line is the standard interstellar radiation field in the solar neighborhood \citep{mathis}, whereas the
dashed line is 
case B radiation field, smoothed to a 64 element resolution. 
The radiation fields have been scaled to the same
integrated flux. Detailed comparison is given in the text.}
\label{comp_ISRF}

\end{figure}

The spectral energy distribution of the radiation field is an important
component in the dust fitting.  Models have been made of interstellar
shocks for velocities relevant for SNR shocks \citep[e.g.][]{shull}. 
They predict a strong UV flux at the shock front with some emission in
the optical as well.  However, as pointed out by
\citet{hollenbachmckee79}, most of the UV flux will quickly be
attenuated in a dense environment.  Once the column density behind the
shock is sufficiently large that molecular hydrogen can reform, the UV
radiation is absorbed and the majority of the radiation will
thus appear in the form of hydrogen lines.  The radiation can then be
modeled through the Case B recombination case \citep{osterbrock}.  The
relative strength of the lines are adopted from \citet{osterbrock},
assuming a temperature of $10^4$ K and an electron density of
$10^4~cm^{-3}$.  
The hydrogen recombination spectrum is relatively soft, lacking the short wavelength emission compared to the local ISRF. 
Figure~\ref{comp_ISRF} shows the different radiation
fields, scaled to the same total intensity.  For computational reasons the
case B radiation field is smoothed to a continuum. The absorption
coefficients behave smoothly as a function of wavelength in this regime and
the smoothed spectrum is sufficient to characterize the radiation field. 

\begin{figure}
\includegraphics[width=10cm]{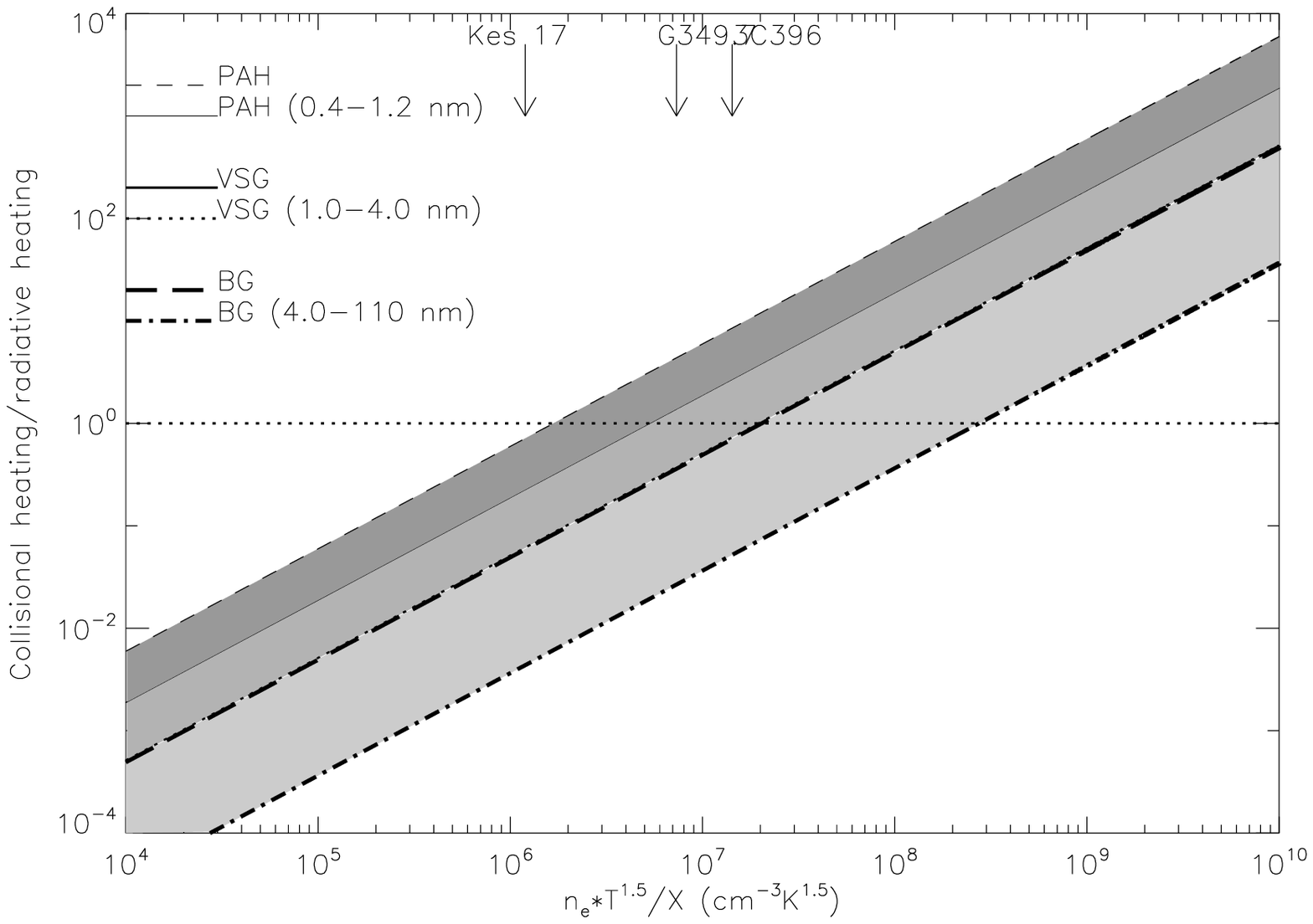}
\caption{Comparison of the ratio of collisional heating to radiative heating. The
abscissa is the quantity $n_eT^\frac{3}{2}/X$, which, for a given grain
size determines whether collisional or radiative heating is dominant. The
radiation field employed is for case B radiation and the strength is adopted
from the DUSTEM fitting. 
The 6 lines are the locus for the three dust species, PAHs, VSGs, and BGs
and the grey shading is darkest for PAHs and lightest for BGs. Note that the
size of the largest PAHs and the smallest VSGs is the same and that the two
lines overlap in the plot. Each species has two lines associated with it;
the largest and smallest particles in the size distribution. The dashed
line indicates equal contribution from radiative and collisional heating. 
The radiation field for the SNRs is found to be higher than that in the interstellar medium.
Representative values for three SNRs are marked. }
\label{heat_comp}
\end{figure}

The equation for collisional heating is derived in \citet{dwekreview}, and is given by: 
\begin{equation}
H_{coll} = \pi a^2n_e\sqrt{\frac{8kT}{\pi m}}\frac{e^{-S^2}}{2S}\int_0^\infty \sqrt{\epsilon}e^{-\epsilon}~sinh(2S\sqrt{\epsilon})~kT~\xi d\epsilon, 
\label{coll_heating}
\end{equation}
where $S=\frac{m}{2kT}v_{gr}^2$, $\epsilon=\frac{mv^2}{2kT}$, and $\xi$ is the efficiency of energy transfer 
from the electron to the grain, assumed here to be 1 \citep{dwekreview}. 
The total heating rate by collisions for a dust grain of radius $a$ is given by 
\begin{equation}
H_{coll}=5.38\times10^{-18}n_ea^2T^\frac{3}{2} ~ erg~s^{-1},  
\end{equation}
where the density is in $cm^{-3}$ and the dust grain size in $\mu$m. The
relative importance of collisional to radiative heating is dependent on the
size of the grain; the larger the grain the more important radiative
heating is.  Figure~\ref{heat_comp} shows the relative importance of
collisional heating to radiative heating assuming a standard interstellar
radiation field as a function of the product $n_eT^\frac{3}{2}$.  The value
of $n_eT^\frac{3}{2}/X$ for the interacting SNRs analysed in
\citet{hewitt} are shown, where the value of "X" is adopted from
Table~\ref{DUST_results}.

Figure~\ref{heat_comp} shows that radiative heating is dominant for BG.
The situation is more complicated for the PAHs where in some cases, e.g. 3C396,
the contribution from collisional heating may be significant.
The PAH or VSG could be heated by collisional heating as well as the radiative heating.   
If there is a strong contribution from collisional heating, we will overestimate the PAH component 
in our fits relative to the VSG and BG components. 
The dominance of
collision vs. radiation heating  depends on exact local
conditions of electron density, temperature and efficiency of energy
transfer from the electron to the grain. In our estimates,
a number of assumptions are used. One assumption is that electron
density is the same as electron density derived  from the Fe lines which is observed with the narrow  IRS slit
while dust is covered by the wider  MIPS SED slit. 
Other uncertainties are local X-ray temperature, density and
ionization rate. 

Moreover, many of sample SNRs show center-filled X-ray morphology
(Combi et al. 2010; Yusef-Zadeh et al. 2003; Harrus \& Slane 1999; see
references in Hewitt et al. 2009; Rho \& Petre 1998) whereas our
infrared images show shell-like morphology. Therefore, it is less
likely that collisional heating from the X-ray emitting  gas is important since the SNR is bordering upon the nearby molecular cloud as
we have detected H$_2$ emission for most sample SNRs.  
An
exception is G11.2-0.3 where both the X-ray and infrared morphology are
shell-like. 
For the case of G11.2-0.3, a
strong correlation between X-ray and IR brightness, a poor fit of the dust continuum by
the DUSTEM model, the presence of fast shocks indicated by the Ne line
ratios and its young age ($\sim$ 2000 yr, inferred from the pulsar),
may indicate  a picture where multiple  physical processes are present. 
This would complicate the determination of the  dominant heating
mechanism. More accurate estimation for individual cases
should be examined with a combined study of infrared and X-ray data. 
This is
out of the scope of this paper. 


\subsection{Modifications to the DUSTEM code}
We have used a beta version of the DUSTEM model by \citet{compiegne} to fit the dust
emission from the SNRs. The model is an update of the dust model by
\citet{desert}. It has been expanded by \citet{bernard} and
\citet{compiegne11}, where the model is described in more detail. The model
and the fitting routines have been modified in several ways for the purpose of this study. 
Note that some of this modifications are now standard
in the publically available version of the code.
 Instead of
only handling a scaling of the local ISRF, now any spectral shape of the radiation
field can be used over the wavelength range 0.09--1.6 $\mu$m.  
 We also  modified
the code to handle the collisional heating only.  
The electron densities and temperatures determined from the ionic lines are not sufficiently high to be able to reproduce the observed SEDs. 
We conclude that the model with only the collisional heating is not  appropriate for this study. 
Unfortunately the current code can not simultaneously handle both radiative and 
collisional heating. We will defer this as future work.
We use radiative heating as the dominant heating mechanism in SNRs for further fits and discussion
in this paper.


The fitting of the SEDs is done using the method described in
\citet{bernard}.  Each of the dust species and the strength of the
radiation field are taken as free parameters in the fit.  This gives a
total of 4 free parameters in the fit. 
We are mainly interested in the relative abundances of the different
dust species and to compare with the general ISM.  We have therefore
re--fitted some of the reference SEDs presented in \citet{bernard} and
\citet{dwek97} using the modified code. 
The resulting abundances of the LMC and the Milky Way agreed well with those previously published. 

\begin{figure}
\includegraphics[width=15cm]{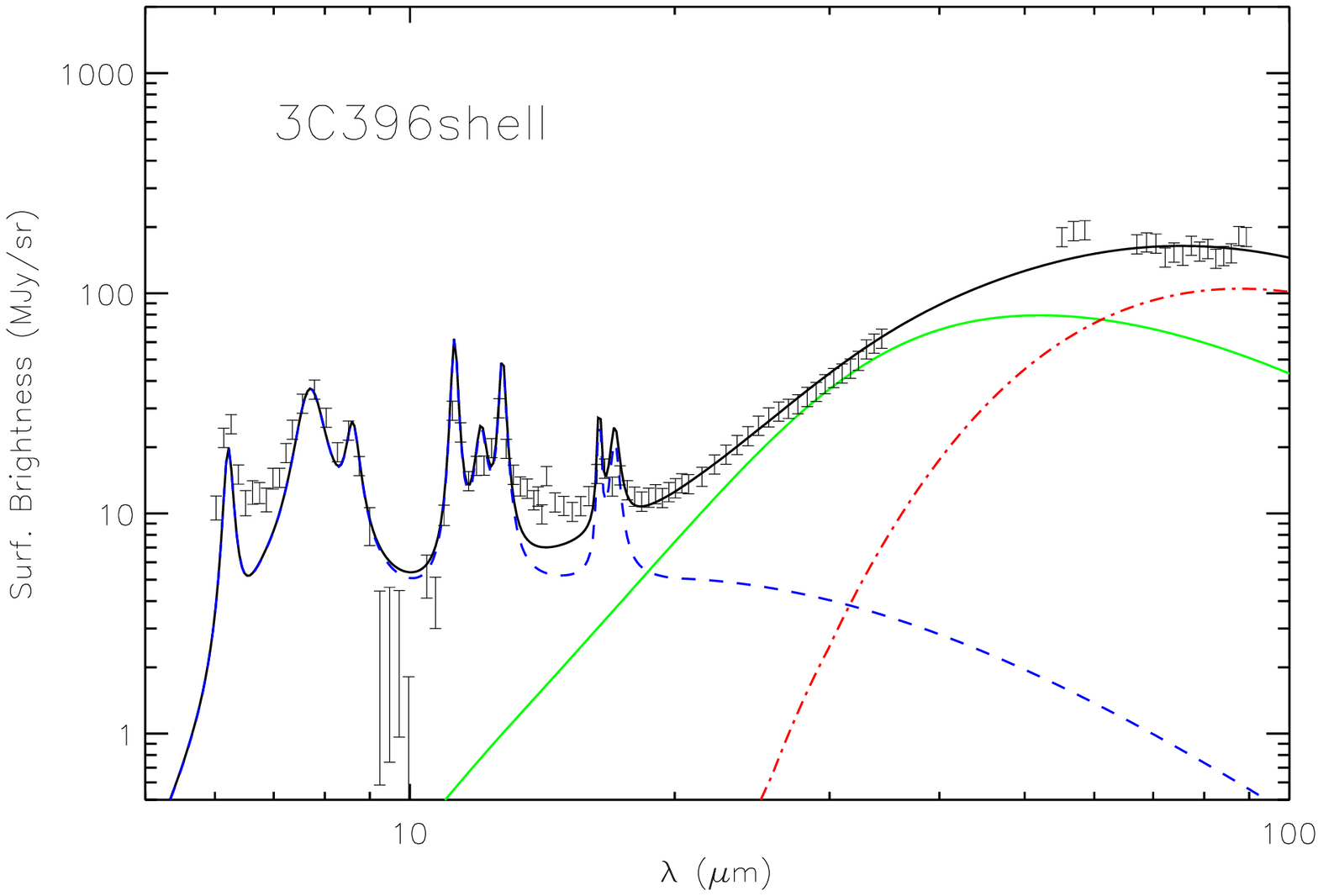}
\caption{The DUSTEM fit for 3C396shell. The spectrum has been de--reddened based on
the foreground hydrogen column density and emission lines have been removed
by Gauss fitting.  The PAH contribution is shown as the blue line, the VSG
contribution as the green line, and the BG contribution is shown as the red line. 
The total fit is shown as the continuous black line. The values for the
best fit parameters are given in Table~\ref{DUST_results}.}
\label{SED_fit_kes17}
\end{figure}

\newpage

\begin{figure}
\includegraphics[width=13.4cm]{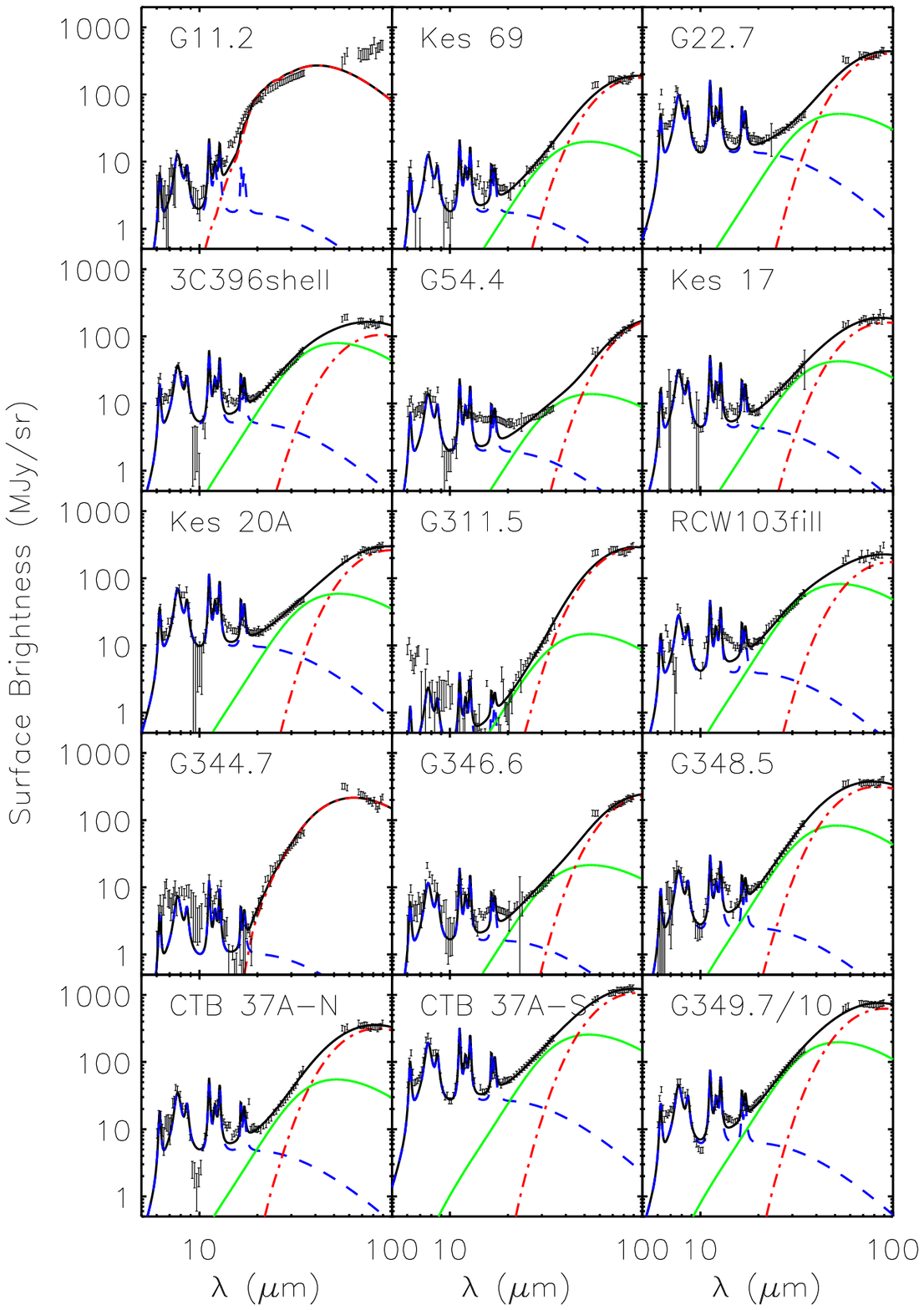}
\caption{The DUSTEM fit for each SNR adopting a case B type  radiation field. The
spectra have been de--reddened and atomic and molecular emission lines have
been removed by Gauss fitting. The PAH contribution is shown as the blue
line, the VSG contribution by the green line, and the BG contribution is
shown as the red line.  The total fit is shown as the continuous black line.} 
\label{SED_fits}
\end{figure}

\subsection{Dust Spectral Fitting and Characteristics of Dust Continuum}
The DUSTEM model assumes three dust species, PAHs, small carbon grains, or
VSGs, and larger silicates, or BGs.  The size ranges of the dust species
are 0.4--1nm, 1--4nm, and 4--110nm for the PAHs, VSGs, and BGs,
respectively. 
Compared to the original description in \cite{desert}, the PAH optical properties were
modified to match the ISO spectrum of the diffuse MW ISM \citep[see][]{compiegne}. 

We fit the mass abundances of  each dust species (Y$_{PAH}$,
Y$_{VSG}$,  Y$_{BG}$) as well as the strength of the radiation field
$F(\lambda)$ (X$_{F(\lambda)}$) for Case B as discussed above. The
abundance is the dust mass to gas mass ratio assuming a hydrogen column of
$N_H=10^{20} cm^{-2}$, which is the expected column density for a molecular gas
density of $10^4~cm^{-3}$ and a shock width of $10^{16}~cm$.  

 For all three dust species, a MRN power-law size
distribution is assumed \citep{MRN}. The slopes and size ranges are the same as provided in the updated DUSTEM code as described in \citet{compiegne}. 
The radiation field has been changed from a
standard ISRF to the shape predicted from Case B recombination. 
Thus, each fit has four free parameters: the 3 dust species and the
strength of the radiation field.  
We tried to fit the spectra by changing the slopes of grain size but we
concluded that our current data are not sensitive enough  to measure the slope. The
slope of the long wavelength part of the SED is very sensitive to the
radiation field and  BG abundances and the spectra between 30-40 $\mu$m is
sensitive to VSG abundances. For a given radiation field, the errors of 
each of the three dust abundances is typically on the order of 20\%.
Uncertainties in the background subtraction amount are $\sim$10\%\ and are
included in the error estimate above. 

Figure~\ref{SED_fit_kes17} shows the fit to the shell of 3C396 in detail and
Figure~\ref{SED_fits} shows DUSTEM fits to all the observed SNRs. A
reasonable fit was obtained in most of cases with a few notable exceptions
such as G11.2-0.3 where the long wavelength part of the fit
shows significant departure from the SED data and for G54.4-0.3 where there is a discrepency at 12--20$\mu$m. The fit results are listed in
Table.~\ref{DUST_results} where the fits for the MW and the LMC are also
provided \citep{bernard}. 
The reduced $\chi^2$  ranges between  3-12.
Systematically higher residuals between the models and spectra are shown between 15 and 20
$\mu$m, resulting in higher reduced $\chi^2$. This is likely  due to the model being relatively crude in modeling the  PAHs.

Our primary conclusions from our spectral fitting are the following. A dust model composed of PAHs, VSGs and BGs provides a reasonable fit to the SEDs of the SNRs. All the SNRs show evidence for PAH emission.
Typically the radiation field is 10-100 times larger than the solar neighborhood interstellar radiation field. The strength of the radiation field is consistent with being created from the shock.
Two SNRs, G11.2-0.3, and G344.7-0.1 show no to little evidence for VSGs and a lower PAH/BG ratio than observed in the diffuse interstellar medium and the LMC. The lack of VSG emission for these two young SNRs is in agreement with the results for CasA (Rho et al. 2008). However, we note that for G11.2-0.3 the fit is not very good and it is thus not clear if the model is applicable in this case


 \subsection{Individual Supernovae: Dust fitting and Shock Parameters}
We discuss the physical parameters for the SNRs that were not presented in \citet{hewitt}. 
The characteristics of the shocks discussed are based on the ionic lines. 

{\bf G11.2-0.3} The morphology of the radio emission from the SNR is a clumpy shell which
is also seen in the MIPS image with a diameter of 4\arcmin . 
\citet{green88} argues the SNR is young and at a distance of $\sim$5 kpc
based of the HI data by \citet{becker}. \citet{green88} favors  the short
distance due to the youth of the SNR and the large physical size the SNR
would have if it was placed at the far distance of $>$26 kpc. A pulsar
associated with G11.2-0.3 is found with a period of 65 ms
\citep{torii97} and an age of $\lesssim$ 2000 yr \citep{kaspi01}. The
hydrogen column towards G11.2-0.3 was found by {\it Chandra} observations
to be 2$\times10^{22}cm^{-2}$ \citep{roberts}.


For G11.2-0.3, the fit to the MIPS SED is very poor.  The reasons for the
discrepancy between the model and observations are not entirely clear,
especially since the fit to the shorter wavelength observations is
relatively good, of similar quality as for most of the other SNRs. It is
worth noting that the off position for the MIPS SED was adopted from
another set of observations for this SNR. In addition, another plausible
possibility is that there is some cold dust along the line of sight that is
influencing the long wavelengths.  Since the SNR is located only 11 degrees
from the Galactic center, the line of sight is rather confused.  Future
longer wavelength observations will help decipher if the latter is the
case.  It is evident from the data already at hand, however, that we see
the 20 $\mu$m silicate feature in emission for this SNR.  There are similar
hints for silicate emission in G344.7-0.1. This is to our
knowledge the first time silicates are seen in emission in an interacting
SNR.

The IRS spectrum shows a rich spectrum of H$_2$ lines, and ionic lines,
particularly iron lines. 
 Utilizing the low excitation lines we find
that a J shock into a pre--shock medium of $10^{3-4}$ $cm^{-3}$ and a
shock velocity of $\sim50-80$ $km s^{-1}$ \citep{hm89} reproduce the majority of the
ionic lines measured.  The exception is the [Fe~II] line at 24 $\mu$m that
is brighter than predicted.  

The [Ne~III] to [Ne~II] ratio is very high, 2.7, suggesting there is in
addition a very high velocity shock ($\ge$ 400$km s^{-1}$) propagating into
a low--density medium, 100 $cm^{-3}$ or less. The shock model comparison
for the [Ne~III] to [Ne~II] ratio is based on the models of
\citet{hartigan} and \citet{mckee87}, and the ratio dependency of the shock
velocity is shown in Figure 6a of \cite{rho01}. We continue to use the same
shock models for comparison with the observed line brightnesses below.

{\bf G22.7-0.2} 
Relatively little is known about the partial shell SNR G22.7-0.2.  The
distance has been estimated with the $\Sigma-D$ relation to be $\sim$3.7
kpc \citep{case}.  At the same distance, a molecular cloud has been
identified in CO observations. We have used CO data \citep{dame01} to
examine the distance and the line of sight column density.  Integrating the
CO emission up the velocity of the molecular cloud at 3.7 kpc provides a
hydrogen column  of 7.8$\times10^{22}~cm^{-2}$.  However, due to the very
large hydrogen column compared with the distance of the SNR and based on
the relatively poor fit with DUSTEM using the value of
7.8$\times10^{22}~cm^{-2}$, we have experimented with PAHFIT \citep{smith07}
to estimate the extinction.  We find that the best fit column density is
high, $\sim$9$\times10^{22}~cm^{-2}$.  However, we also find that a fit that
is almost as good can be obtained with a lower hydrogen column.  A hydrogen
column as low as 1.5$\times10^{22}~cm^{-2}$ provides a fit that is only 10\%
worse in a chi square sense than the best fit value.  We therefore use the
extinction value of 7.8$\times10^{22}~cm^{-2}$ determined directly from the
CO measurements. 

There is no detection of [Ne~III] in G22.7-0.2. The lines of [Ne~II], 
[Fe~II] at 26 $\mu$m and  [Si~II] are detected. Comparison with shock models
implies a shock of $\sim$ 90 $km s^{-1}$ propagating into a medium with a
density of $10^3 cm^{-3}$.  

{\bf G311.5-0.3} G311.5-0.3 is a relatively small (4\arcmin\ diameter) shell SNR.  The
distance has been estimated by the $\Sigma-D$ relation to be 12 kpc,
whereas emission from CO data has given a radial velocity of $v=+39.7~km~s^{-1}$
which gives a distance of 14.8 kpc. The detection of H$_2$ emission implies
an interaction with molecular clouds so we have adopted the distance of
14.8 kpc for this study. The total hydrogen column density is
2.5$\times10^{22}~cm^{-2}$ obtained from CO data\citep{dame01}. To our
knowledge, there has been no detection of OH masers or other
indications of interaction with a molecular cloud so our H$_2$ line
detection is likely the first evidence of such an interaction. 

The ionic lines suggest a moderately fast shock, 40--90 $km s^{-1}$
propagating into a medium with an initial density of $10^{4}$ $cm^{-3}$. 
The [Ne~III] to [Ne~II] ratio points
towards either a shock of 100 $km s^{-1}$ with an electron density of $10^3$
$cm^{-3}$ or a faster shock, $\sim170$ $km~s^{-1}$ and an electron density
of 100 $cm^{-3}$. 

{\bf RCW 103} RCW 103 is a well studied young ($10^3~yr$) shell morphology SNR.  The
distance has been estimated through HI absorption to be 3.3 kpc, based
on a velocity of $-56$ km $s^{-1}$ \citep{caswell}.  The hydrogen column has
been estimated both through X--rays \citep{gotthelf} and near--infrared
spectroscopy \citep{oliva90} and has been found to be
7$\times10^{21}~cm^{-2}$.  Previous near--infrared and ISO observations
have identified H$_2$ lines slightly outside the SNR, indicating the
material has been heated by soft X--rays \citep{oliva90,oliva99}.  

Due to the size of RCW 103 and the extended emission from the shock, the
local off positions for the short low observations are located within
the shocked material.  There is little dust continuum associated with
the parts of the shock that are used as sky background.  However, there
are emission lines present from the shock which results in an
over subtraction of these lines in the on position.  We have masked out
the regions where the over subtraction have produced negative residuals
in the continuum spectrum. 

The ionic lines suggest a shock propagating with a velocity of some 
80 $km s^{-1}$ into a pre-density of $10^4$ $cm^{-3}$.  However, the [O~I] line
is too weak compared to that predicted by \citet{hm89} for a velocity of
80 $km~s^{-1}$ and a density of $\sim 10^{4} cm^{-3}$. The model predicts a
surface brightness of $\sim 10^{-2.3} ergs^{-1}~cm^{-2}~sr^{-1}$. Although a
small amount of [O~I] emission can have been subtracted with the diffuse
sky emission, it cannot explain this large difference. The current
determination of the [O~I] surface brightness flux is in good agreement
with  the measurements by \citet{oliva99}. The [Ne~III] to [Ne~II] line
ratio of 0.37 suggest either a shock with a velocity of 100 $km~s^{-1}$ and
an electron density of 1000 $cm^{-3}$ or a faster shock, 210 $km~s^{-1}$ and
$n_e=100$ $cm^{-3}$.

{\bf Kes 20A} Little is known about the SNR Kes 20A. \citet{whiteoak} identified a well
defined  eastern arc of the SNR and a weaker western arc.  Based on CO line
data, we have estimated a radial velocity of +30 $km~s^{-1}$, resulting in a
distance of 13.7 kpc.  We have found no estimate of the foreground hydrogen
column in the literature. Instead we have used CO observations as for G22.7-0.2.  
By integrating the emission up to the velocity of the
SNR, we estimate a total H$_2$ column of 4.19$\times$ 10$^{22}$ cm$^{-2}$,
which corresponds to a hydrogen column of 8.4$\times$10$^{22}$ cm$^{-2}$.
The shortest wavelength slit does not extend far enough to completely
extend to a spatial region devoid of radio emission. Thus, the whole slit
is likely within the shock. However, comparing with the IRAC images, it is
evident that there is little emission and the sky position is a good estimate of
the background emission. Kes 20A shows a very limited number of ionic
lines. 
 Only [SiII] (34.82 $\mu$m) and [Fe~II] (35.35 $\mu$m) are
detected.  Both lines indicate a relatively low velocity shock ($V_s\sim35-70~km~s^{-1}$) into a low
density medium ( $n_0\sim10^3~cm^{-3}$). 

{\bf G54.4-0.3} G54.3-0.3 is a large shell--like SNR in a complex region.  The distance has
been estimated to 3 kpc based on the detection of a molecular cloud and the
association of the SNR with OB associations and HII regions
\citep{junkes92}.  The hydrogen column has been estimated from {\it ROSAT}
X--ray data and is found to be $10^{22}~cm^{-2}$\citep{junkes96}. 
There has been previous identification of a molecular cloud in the vicinity
of G54.3-0.3  
but direct evidence for interaction of the SNR with surrounding molecular
material was not shown. 

G54.4-0.3 is another SNR where only few  ionic lines are observed. 
The [Fe~II] lines at 26 and 35 $\mu$m suggest a shock into a low density
medium ($10^3~cm^{-3}$), which is confirmed by the  detection of [SiII]. 
The shock velocity determined from the 3 lines is in the range 40--80 $km~s^{-1}$.

{\bf G344.7-0.1} SNR G344.7-0.1 is an asymmetric shell SNR in the radio. The main
features of the shell can be seen  in the MIPSGAL image except for the
western direction where there is little enhanced emission at 24 $\mu$m. 
The distance has been estimated from the $\Sigma-D$ relation to be 14
kpc \citep{dubner}.  Based on ASCA X--ray observations the hydrogen
column was estimated to be 5$\times10^{22}~cm^{-2}$ \citep{yamauchi}.  No OH masers have been
associated with this SNR \citep{green97}. 


The [Ne~III] to [Ne~II] line ratio of 0.54 suggests a high velocity shock of
$\sim280~km~s^{-1}$ and an electron density of 100 $cm^{-3}$.  The ionic
lines further suggest that the pre-density material had a density of some
$10^4~cm^{-3}$ and that the shock velocity is some 80 $km~s^{-1}$. The
models of higher density shocks predict stronger [O~I] and [SiII] lines than
observed. 

{\bf CTB 37A} CTB 37A is a poorly defined shell SNR overlapping with
G348.5-0.0 and CTB 37B.  Its distance has been constrained to be between
8 and 11 kpc.  We have chosen the distance of 11 kpc that is associated
with the HI emission feature.  Chandra X--ray observations have
determined a hydrogen column of 3.2$\times$10$^{22}$ cm$^{-2}$
\citep{aharonian}.  \citet{frail} detected several OH masers associated
with the shell of CTB 37A. 

The region is very complex and the SNR covers a large area on the sky. 
The LL2 does not extend completely outside the radio
emission of the SNR.  
However, the IRAC and MIPS 24 $\mu$m images show negligible emission 
in the outer regions covered by the IRS long slit.  
We have compared with the IRAC images and the 24 $\mu$m image
that there is little to no excess continuum emission in the outer parts and 
 we have used that outer part as our sky position. 

For CTB 37A-N,  the fit is mediocre at the 9.5 $\mu$m range.
This is in contrast to
CTB 37A-S where the short wavelength range is fit relatively well. A likely
explanation is uncertainties in the dereddening of the source.  For CTB37A,
one reddening value is adopted for the whole SNR, despite the fact that it
covers 15\arcmin radius area close to the Galactic center.  It is thus
likely the reddening can vary across the SNR due to unrelated foreground
material. Since the silicate absorption feature is a strong function of the
extinction compared to the rest of the mid--IR extinction curve, the fit
can be substantially improved by increasing the extinction.  However, in
order to provide a much better fit, the hydrogen column has to be almost
doubled compared to the value determined from X--rays.  Even altering the
hydrogen column to $5.5\times10^{22}~cm^{-2}$ only changes the reduced
$\chi^2$ by $\sim 20\%$.  The ratio of dust species remain almost the same.
 The quality of the fit improves a bit, the reduced $\chi^2$ is 3.5
compared to 5 adopting the X--ray determined hydrogen column. The
subsequent analysis is based on the best fit adopting the hydrogen column
determined from the X--ray. The conclusions do not change significantly
adopting the higher hydrogen column.

We find the shock characteristics to be slightly different at the two
locations observed within CTB 37A. For the CTB 37A-N position we find a
density of $10^{3.5}~cm^{-3}$ and a shock velocity of $\sim 100~km~s^{-1}$
whereas the CTB 37A-S region appears a bit denser, $10^{4}~cm^{-3}$ and the
shock is slower, $\sim 75 km~s^{-1}$. The observed [O~I] line is too weak for
such dense shocks where a surface brightness of $\sim 10^{-2}
erg~s^{-1}~cm^{-2}~sr^{-1}$ is predicted. 





\section{Discussion}
We present the results for the DUSTEM fitting to the SEDs of the SNRs. The
ratio of the dust species is determined and is compared with the same
ratios for Galactic plane dust and dust in the LMC interstellar medium. We
further discuss the results in view of dust destruction models. 
\subsection{Dust Properties of SNRs}

\begin{figure}
\includegraphics[width=10cm]{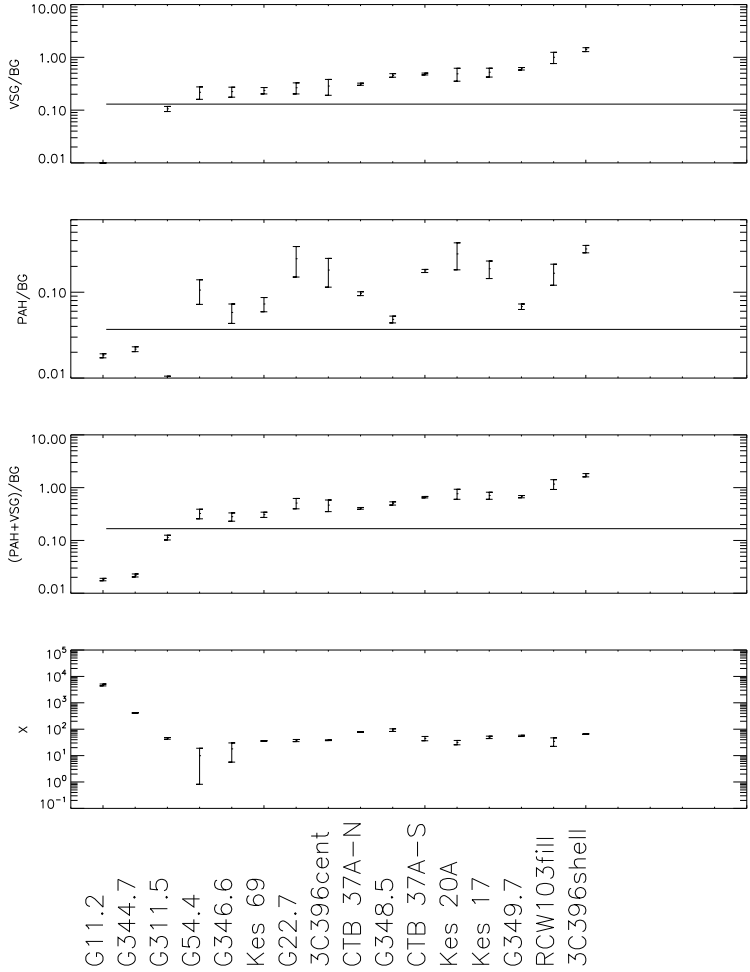}
\caption{From top to bottom: The ratio of VSGs, PAHs, and PAH+VSGs to BGs, assuming a Case B radiation field. 
The ratios determined for the plane of the Milky Way  are shown as the solid line in each plot. 
Most of the dust ratios are above the ratio determined in the MW plane at least at a 2$\sigma$ level. However,  a small group of SNRs have  ratios of carbon dust to silicates smaller than the MW plane ratio. These are associated with a strong radiation field.  }
\label{plot_ratio_caseb}
\end{figure}

We present the fitting results using Case B radiation field in Table
~\ref{DUST_results} and the results are further summarized in
Figure~\ref{plot_ratio_caseb}. The radiation field in SNRs is higher than
those of ISM and the average value of the Milky way and LMC; this conclusion is
independent of the spectral shape of the radiation field. Most of the SNRs are
well fit with a radiation field of 50--200 times the local interstellar
radiation field, in general agreement with the modified black body fits
performed above.  
 Although the ISRF is expected to be stronger in the inner
parts of the Galaxy where most of the known SNRs reside, we do not expect this to
be able to account for the high radiation field.  Typically, closer to the
Galactic center, the ISRF is only a few times the strength of the local
ISRF.  Instead, we suggest below that the enhanced radiation can be explained by
hydrogen recombination. Note that since the absolute gas column density of
the SNR shock front is not known, we cannot compare the absolute abundances
between SNRs or the Milky Way. However, the ratios can directly be compared
between different SNRs and other environments, e.g. the interstellar
medium. 

\begin{figure*}[!h]
\plotone{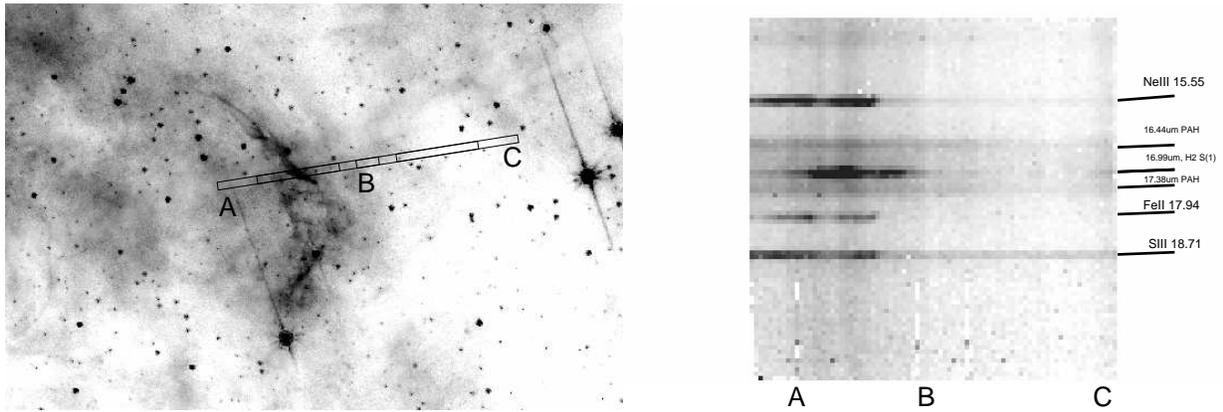}
\caption{ Left: a {\it Spitzer} GLIMPSE 8 $\mu$m image of Kes 17. Overlaid is the four slit positions of the LL2 module. 
Right: The combined 2 dimensional spectrum from the four LL2 sub slits. 
Enhanced PAH emission from 15--20$\mu$m is seen at the
position of SNR shell (A to B on slit) indicating that PAH emission is
associated with the SNR Kes 17.  }
\label{kes17_slit}
\end{figure*}

\begin{figure*}[!h]

\plotone{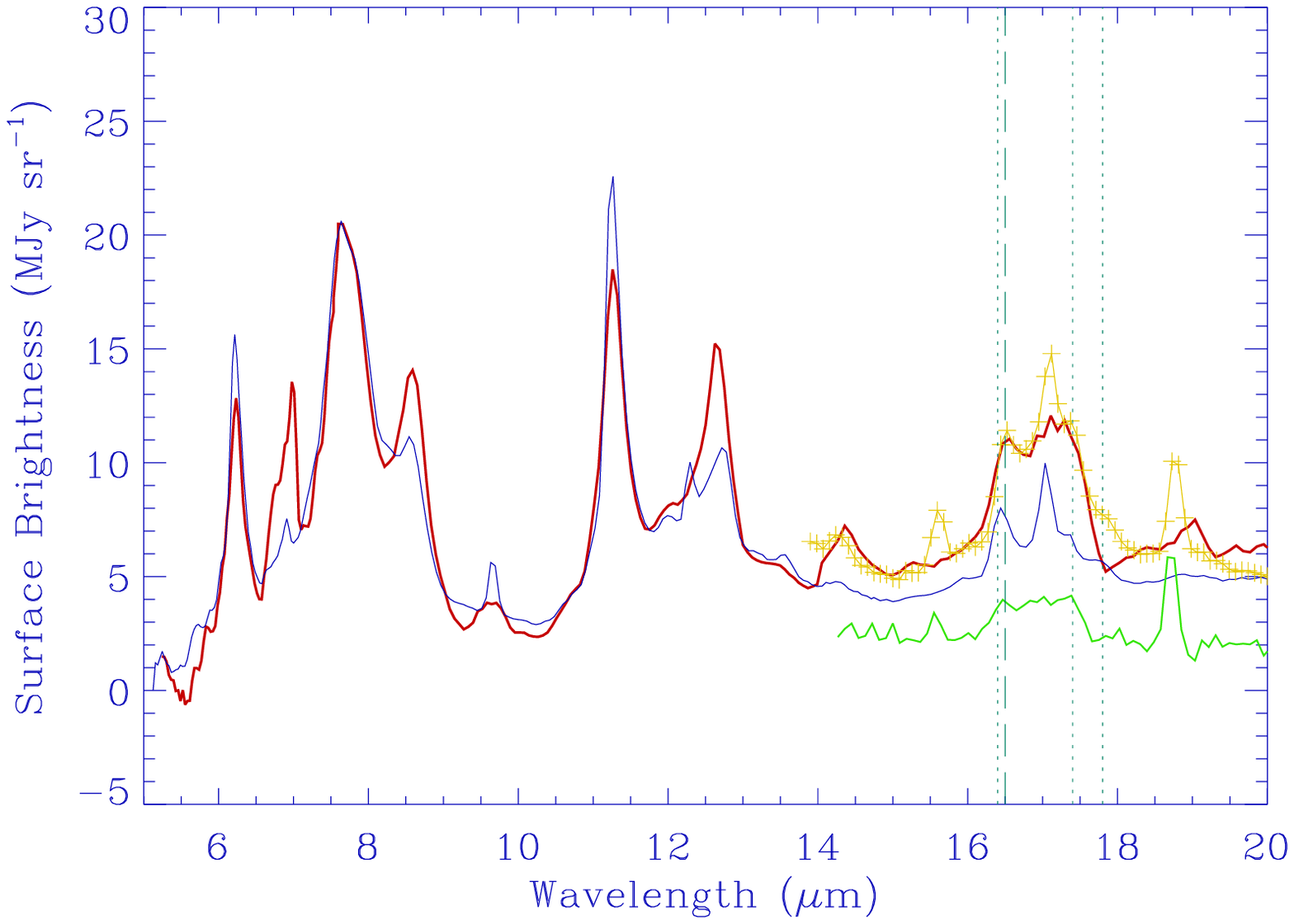}
\caption{ Comparison of PAH emission in the SNR Kes 17 (red curve) with those of LkH$\alpha$ 234 
(PDR: blue)  and NGC7331 (yellow). The 15-20 $\mu$m PAH emission in the background of Kes 17 (green) is also compared,
showing that the plateau bump is weaker than that of Kes 17 and that the shape was somewhat altered.
 The shape of the PAH emission between 5-14 $\mu$m in Kes 17 is almost identical to that in 
the Galactic background, suggesting little processing of the smallest grains. }
\label{PAHcomp}
\end{figure*}

The dust spectral fitting indicates the presence of PAH emission in most of SNRs 
as shown in Figure \ref{PAHcomp}.
The dip centered at 9.7 $\mu$m could indicate either presence of
PAH emission or high extinction from silicates. 
However, the feature is present after the foreground extinction correction of the  SNRs 
and is thus likely due to PAH emission. 
We here show a detailed example of PAH emission using an example of the SNR Kes 17.
Figure \ref{kes17_slit} shows the location of the slits covering Kes 17.  
The "A" location is within the SNR, and the location between the B and C is the surrounding ISM.
The PAH emission around 17 $\mu$m is enhanced at the shell of the SNR Kes 17.
The PAH spectrum of Kes 17 is shown in Figure \ref{PAHcomp} where it is compared with the emission from LkH$_\alpha$ 234 and NGC7331, a PDR and a nearby spiral galaxy with strong PAH emission, respectively. 
The PAH emission is always detected in our background spectra since all the SNRs are located in the Galactic plane. 
However, the PAH emission is clearly enhanced in the shocked region and is present after background subtraction. 
The 15-20 $\mu$m PAH bump in Kes 17 is stronger than that of LkH$\alpha$ 234 whereas
the PAH features between 5-14 $\mu$m are almost identical (the slight
differences are likely due to residuals from line subtraction or 
extinction correction). However, the 15-20 $\mu$m feature of Kes 17 is flatter than
that of the PDR and the feature shows a plateau shape. 
The feature  is similar to that of NGC 7331.
The ratio of the emission from the 15-20 $\mu$m plateau to the 6.2 $\mu$m PAH feature  is 0.4 for Kes 17 and other H$_2$ emitting SNRs,
similar to NGC 7331 but larger than for the PDR. 
This ratio is 
significantly smaller than that of the young SNR N132D \citep{tappe06}.  
The difference between our sample and N132D is the environments of SNRs.
Many of the interacting SNRs are in a  dense molecular cloud
environment with a slow shock, while N132D 
in less dense environment has a strong shock which significantly  
destroy small PAHs. 
We note that whenever we detected PAH emission in SNRs there is PAH emission 
in the background as well (see Figure \ref{kes17_slit}), indicating the 
PAH emission is from shock processed 
dust instead of shock generated dust.
For some of the observed SNRs, one or more of the slits did not cover a background region.  
Full spectral mapping 
would be needed for further detailed studies of PAH processing through shocks.



The three different dust components dominate in different parts of the
observed SEDs. Shortwards of 20 $\mu$m the PAH component dominates the
spectrum. Between 20 $\mu$m and up to 30 $\mu$m the emission from VSGs is
in general the main contribution,  whereas the MIPS SED traces the BGs.
Between 30--35 $\mu$m the VSGs and BGs can both contribute depending on the
strength of the radiation field. For two of the SNRs, G11.2-0.3 and
G344.7-0.1, the SEDs can be fit without the need of VSGs. We discuss these
two SNRs separately.

The shape of the continuum  is well reproduced by the model for each of the sample SNRs except G11.2-0.3 and G344.7-0.1. 
Where the model
in general fails to fit the data in detail is in the shape of the PAH
features at shorter wavelengths. The PAH feature at 12.3 $\mu$m is not well
reproduced. 
 It appears the observations have a red 'wing' that is stronger
than predicted. This is likely not an anomaly with the PAH emission from
the shock heated dust. In similar spectral fitting to the Horsehead Nebula,
\citet{compiegne} also finds that the model underpredicts the emission slightly
in this wavelength range. 

The reduced $\chi^2$ for the fits are in general moderately poor.  A closer
look at the individual fits shows that a large part of the discrepancy is
due to differences in the PAH dominated part of the SEDs.  Here we note the
origin of the discrepancy between the model and the observed PAHs but defer
a detailed discussion of the PAH evolution to a future paper.  We further
note that the reduced $\chi^2$ determined for the SNRs is similar to the
reduced $\chi^2$ for the Galactic plane measurements.  We thus conclude the
model fits to the dust from the SNRs captures the fundamental large scale
features of the dust emission.


The abundance of each of the three dust species for each SNR is given in
Table~\ref{DUST_results}. The SNRs can be placed into two categories.
 The majority of the SNRs have ratios of VSGs to BGs and PAHs to BGs that
are higher than those observed in the diffuse interstellar medium of
Galactic plane. This group includes G22.7-0.2, Kes17, 3C396,  CTB37A-S,
RCW103, kes 20A, and  G349.7+0.2. 
The radiation field derived for this sample is intermediate in strength, much lower than that of the group with a small ratio of carbon to silicate grains. 
 A smaller group has a low ratio of carbon grains to silicates, including G11.2-0.3 and  G344.7-0.1. 
This group is also further characterised by a very high radiation field, sufficiently high to see silicates in emission at 20 $\mu$m. 

The derived PAH abundances are also in general higher than observed in the MW plane. 
The larger error bars for the PAH abundances, due to their general low abundance relative to the VSGs and BGs, makes it difficult to find any correlation between the derived PAH abundances and other parameters. 
However, their overabundance relative to the MW plane support a shattering scenario for dust processing together with the results for the VSGs. 
Note that the two SNRs with low VSG to BG ratios also have a low abundance of PAHs which confirms the shattering scenario.


\subsection{Dust Processing Through Sputtering And Shattering}
The ratios of carbon dust (PAHs and small grains) to silicates can be explained
through dust destruction scenarios.  Two dust destruction
scenarios are sputtering \citep[e.g.][]{dwek96} and shattering
\citep[e.g.][]{borkowski,jones} which are discussed below. 

\subsubsection{Sputtering}
It is predicted that sputtering will alter the grain size
distribution in fast shocks \citep{dwek96}.  The efficiency of sputtering depends on size
of grains and is most efficient on smaller grains resulting in a mass
distribution with a relative deficit of small grains compared to the general ISM grain size distribution. 

We find for a low value of both
the PAH and VSG abundance relative to the BG abundance for G11.2-0.3, G344.7-0.1, and
G311.5-0.3.  In fact, the SED for G344.7-0.1 suggest the presence of no VSGs at all
and only a small amount of PAHs. 
 The situation is less clear for G311.5-0.3
since the emission at all wavelengths is rather weak and the abundances
measured, for all three dust species, is uncertain, due to a complicated background diffuse Galactic background near G311.5-0.3. 
For  G11.2-0.3 and G344.7-0.1, the neon line
measurements suggest a fast shock (280 $km s^{-1}$) into a medium with an
initial density of some 100 $cm^{-3}$.  
Comparing with the calculations
by \citet{dwek96} we see that at these temperatures and densities,
significant amount of sputtering can occur, which thus provides a  likely explanation for
the low abundance of PAHs and VSGs relative to BGs in G11.2-0.3 and G344.7-0.1. 
Note also that for G11.2-0.3, and G344.7-0.1 we see evidence for the silicates in emission at 21 $\mu$m. 
We find sputtering is efficient at high shock velocities and
relatively low densities as observed in G11.2-0.3, and G344.7-0.1 compared to those observed in the other  interacting
SNRs where the sputtering in general do not play a dominant
role as suggested by \citep[e.g.][]{jones}.  
 A more sophisticated dust model combined with an  abundance analysis may be required
to understand the role of sputtering in the two young SNRs.

\subsubsection{Shattering}

The ratio of Y(VSG)/Y(BG) ranges between 0.11-1.4 for the SNRs not showing silicates at 21 $\mu$m in emission . 
The ratio in SNRs is a factor of a few larger than those of the MW (0.13) and the LMC for most of the SNRs.
We interpret the higher ratio is indication of shattering, big grains being destroyed
by shocks turning them into small grains.
\citet{jones} suggest that in dense, relatively slow shocks, grain
processing through shattering will occur. Contrary to the case of
sputtering, large grains are predominately affected and silicates are more
vulnerable than graphite grains \citep{jones}.  For the typical shock
velocities determined for the shocks, shattering is expected to be
efficient for grains larger than some 400--1000\AA\ which will be shattered
and we thus expect shattering to occur for the BGs. 

It is possible that some of the BGs are shattered into the size distribution of the VSGs. 
The ratios of VSG to BG seem to be higher for the high speed shocks. 
However, the standard VSGs are predominantly comprised of carbon and silicates.  
Silicates are characterized by strong emission features at 10 $\mu$m and 20 $\mu$m.  
However, these features are not observed in most of the SNR spectra. 
We have thus examined whether silicate grains with a size range and distribution similar to that of  the VSGs would produce emission features inconsistent with the observed spectra. 
The strength of the silicate emission features depend on the strength of the radiation field. 
A Case B radiation field 500 times stronger than the default strength, the strength of the 10 $\mu$m emission feature is very weak compared to the emission from a similar amount of carbon VSGs (in mass) irradiated by the same radiation field. 
At 20 $\mu$m the emission peak is comparable in strength to the emission from the carbon VSGs. 
Significantly higher radiation fields are required to produce a substantial 10 $\mu$m feature compared to the strength of the PAHs. 
However, the only two SNRs where such a strong radiation field is observed are already fit purely by PAHs and BG silicates and no VSGs. 
It is therefore possible to have a population of silicates towards small sized dust particles and not have them show up as emission features at shorter wavelengths.

Since the BGs are over five times
as abundant than the VSGs in the ISM  (in mass), even a small processing
of BGs into VSGs would increase the ratio of the two substantially.  For
example,  a shattering of 10\% of the BGs into VSGs will increase an
initial VSG to BG ratio of 0.2 to 0.3.  If 50\% of the BGs are processed
to VSGs the ratio will be 1.2, assuming all the shattered fragments end
up as VSGs. 
\citet{jones} calculates the grain size where 50\% of the grains will be
reprocessed for a given shock velocity.  For shocks faster than $\sim$ 100
km/s, the limiting grain size is $\sim$400 \AA.  Almost all grains
larger than the critical size will be destroyed and almost all grains
smaller will survive.  We can thus calculate the predicted amount of large
grains shattered from the input MRN distribution.  Adopting the minimum
and maximum grain size from the DUSTEM code and a critical size of
400\AA, we find that 46\% of the large grains will be destroyed by a
shock faster than 100 km/s.  Thus, the ratio of carbon grains (which are
too small to be affected by shattering) to the large silicates would 
double by the removal of the large grains alone with a further increase
from the smaller particles produced in the shattering process.  A shock
speed of 100 $km s^{-1}$ is where the smallest grains are affected.  At
both lower and higher speeds only larger grains will be completely
shattered. 

The majority of the shock velocities determined above are in the
range 50--80 $km s^{-1}$ and we would thus expect less of the BGs being
shattered than for a 100 $km s^{-1}$ shock, which is indeed what is
observed for most of the SNRs.  In principle we should see a monotonic
dependence of the VSG to BG ratio as a function of shock speed in that,
the slower the shock, the less shattering occurs.  There is no strong correlation seen which is likely to be due to the uncertainty in the derived shock velocities (see above). 
However, there is a tendency for the highest VSG to BG ratios to correspond to the  highest derived shock velocities.  
Further, as shown for a few of the SNRs (e.g. G11.2-0.3, \citet{lee09}, and 3C396, \citet{koo}) the H$_2$ emission seems to be more localized whereas the [Fe~II] emission is a bit more diffuse. 
All SNRs where the shock speed has been
determined to be larger than $60 km s^{-1}$ have an increased VSG to BG
ratio and those with slower shocks tend to have ratios located closer to the 
ratio found in the MW plane.
 The two exceptions are Kes 20A and RCW 103.  Due to
the limited number of ionic lines for Kes 20A, the shock velocity is somewhat uncertain. 


\begin{deluxetable}{lllllllll}
\rotate
\tabletypesize{\scriptsize}
\tablecaption{The parameters obtained for the DUSTEM fits adopting a Case B radiation field. The dust abundances are relative mass fractions to a fiducial hydrogen column density of N$_\mathrm{H}=10^{20}cm^{-2}$. X is the strength of radiation field relative to the case B radiation field 
of which the total energy is normalized by the ISRF.
$\chi^2$ is the reduced  goodness of the fit using DUSTEM model.}

\tablehead{\colhead{ Object} & \colhead{Y(PAH)} & \colhead{Y(VSG)} & \colhead{Y(BG)} & \colhead{Y(TOT)} & \colhead{X} & \colhead{Reduced $\chi^2$} & \colhead{$\frac{Y(PAH)}{Y(BG)}$} & \colhead{$\frac{Y(VSG)}{Y(BG)}$}}
\startdata
MW (plane) & 3.11E-4 & 1.09E-3 & 8.42E-3  & 9.82E-3 & 1.46 & 7.87 & 3.69E-2 & 1.3E-1 \\
MW (diff) & 4.83E-4 & 6.38E-4 & 4.68E-3 & 5.8E-3 & 0.8 & 11.87 & 10.3E-2 & 1.36E-1 \\
LMC & 9.7E-5 & 3.47E-4 & 3.16E-3 & 3.6E-3 & 1.85 & 1.12 & 3.07E-2 & 1.1E-2 \\

G11.2&1.8(0.1)E-5&0.0(0.0)E 0&1.0(0.1)E-3&1.0(0.1)E-3&4.8(0.2)E+3&           12&
1.8(0.1 )E-2&0.0(0.0 )E 0\\
Kes 69&2.4(0.4)E-3&7.9(1.5)E-3&3.3(0.5)E-2&4.4(0.5)E-2&3.6(0.6)E+1&           7&
7.3(1.5 )E-2&2.4(0.6 )E-1\\
G22.7&1.8(0.3)E-2&2.0(0.5)E-2&7.4(1.3)E-2&1.1(0.1)E-1&3.7(0.7)E+1&           9&
2.5(0.6 )E-1&2.7(0.8 )E-1\\
3C396cent&1.3(0.2)E-2&2.1(0.4)E-2&7.2(1.2)E-2&1.1(0.1)E-1&3.8(0.6)E+1&
           9&1.8(0.4 )E-1&2.9(0.7 )E-1\\
3C396shell&3.9(1.2)E-3&1.7(0.6)E-2&1.2(0.3)E-2&3.3(0.7)E-2&6.6(2.1)E+1&
           6&3.2(1.3 )E-1&1.4(0.6 )E 0\\
G54.4&9.9(1.9)E-3&2.0(0.4)E-2&9.3(2.2)E-2&1.2(0.2)E-1&9.9(2.0)E+0&          12&
1.1(0.3 )E-1&2.2(0.7 )E-1\\
Kes 17&4.4(0.9)E-3&1.2(0.3)E-2&2.3(0.5)E-2&4.0(0.6)E-2&4.9(1.0)E+1&           3&
1.9(0.5 )E-1&5.2(1.7 )E-1\\
Kes 20A&1.5(0.3)E-2&2.7(0.7)E-2&5.5(1.3)E-2&9.7(1.5)E-2&3.2(0.7)E+1&           5
&2.8(0.9 )E-1&4.9(1.7 )E-1\\
G311.5&3.6(0.4)E-4&4.7(0.6)E-3&4.4(0.4)E-2&4.9(0.4)E-2&4.5(0.4)E+1&           7&
8.1(1.1 )E-3&1.1(0.2 )E-1\\
RCW103fill&5.6(1.6)E-3&3.4(1.1)E-2&3.4(1.0)E-2&7.3(1.4)E-2&3.5(1.0)E+1&
           8&1.7(0.7 )E-1&1.0(0.4 )E 0\\
G344.7&1.2(0.1)E-4&5.7(0.0)E-9&5.7(0.2)E-3&5.8(0.2)E-3&4.1(0.1)E+2&           6&
2.2(0.1 )E-2&1.0(0.0 )E-6\\
G346.6&4.5(0.8)E-3&1.7(0.4)E-2&7.7(1.6)E-2&9.8(1.7)E-2&1.8(0.3)E+1&           8&
5.8(1.6 )E-2&2.2(0.7 )E-1\\
G348.5&1.3(0.2)E-3&1.3(0.2)E-2&2.8(0.3)E-2&4.2(0.4)E-2&9.3(1.2)E+1&           5&
4.8(0.8 )E-2&4.6(0.9 )E-1\\
CTB 37A-N&3.1(0.4)E-3&9.9(1.8)E-3&3.2(0.4)E-2&4.5(0.4)E-2&7.8(1.0)E+1&
           6&9.6(1.6 )E-2&3.1(0.7 )E-1\\
CTB 37A-S&3.0(0.6)E-2&8.2(1.9)E-2&1.7(0.3)E-1&2.8(0.4)E-1&4.5(0.9)E+1&
           5&1.8(0.5 )E-1&4.8(1.4 )E-1\\
G349.7&5.6(0.9)E-2&5.0(1.0)E-1&8.2(1.3)E-1&1.4(0.2)E 0&5.6(0.9)E+1&           6&
6.8(1.6 )E-2&6.1(1.5 )E-1\\
\enddata
\label{DUST_results}
\end{deluxetable}

\subsection{Integrated Dust Mass}
With an estimate of the dust mass within the IRS and MIPS SED beam we
can derive the total dust mass in each SNR by assuming the 24 $\mu$m and 70 $\mu$m emission trace the total dust mass.  We have integrated the
total emission from the SNRs from both the MIPS 24 $\mu$m and  70 $\mu$m
images.  Even though the derived 70$\mu$m fluxes are in the nonlinear
regime we can still estimate the total flux from the SNR.  
 This approach works best for relatively small, well defined
SNRs.  Examples are G11.2-0.3, G311.5-0.3, G344.7-0.1, and Kes 17 where we can
easily trace the emission and where the region is not confused by e.g.
HII regions and other bright  unrelated objects. 
We note that for many of the SNRs the total dust mass is a lower limit for the following reason. 
For the extended SNRs we have used the radio contours in conjunction with a  with a visual inspection of the images to identify the regions with dust emission associated with the SNR. 
We have thus excluded regions in the SNRs close to very bright stars or to regions that appears to be HII regions along the line of sight to the SNR, as is for example the case of CTB 37 near the south position. 
We therefore will be missing some of the faint dust emission from the SNRs, especially for the large area SNRs as CTB 37, and 3C396. 
In addition, the central position in 3C396 is located near a very bright star and is not used for the estimate of the total dust mass within the SNR.

Moderately bright stars within the SNR in both the 24 $\mu$m and 70 $\mu$m
images have been masked before the flux has been integrated across the
SNR.  The background level has been estimated from adjacent regions of
clear sky and has been subtracted from the emission form the SNR.  It is
evident from the IRS spectra and the MIPS SED that the broad band
emission is contaminated by line emission. In particular the iron line
at 26 $\mu$m and the [O~I] line at 63 $\mu$m are strong.  However, since we
have assumed similar properties for the gas and dust within the MIPS SED
and IRS beam as for the whole SNR, this fractional contribution will be
the same in both cases. Table~\ref{int_dust} shows the correction factor
(one over the fraction of the SNR emission covered by the MIPS SED )  at 24
and 70 $\mu$m, and the total mass for each the the dust species as well as
the total shocked dust mass for the SNR. The total mass was derived using
the average of the two correction factors. 
\begin{table}[ccccccc]
\caption{Integrated dust masses for the SNRs. The total flux associated with each SNR in the 24 and 70 $\mu$m images have been calculated and the dust mass determined from the SED fitting has been scaled accordingly.}
\begin{tabular}{lllllll}
\hline
SNR & Cor factor (24 $\mu$m)  & Cor factor (70 $\mu$m) & PAH mass & VSG mass & BG mass & Total mass\\
& & & M$_\odot$ & M$_\odot$ & M$_\odot$ & M$_\odot$ \\
\hline
   G11.2 & 2.3e+01& 3.8e+01& 1.6e-04& 0.0e+00& 8.6e-03& 8.8e-03\\
   Kes69 & 9.9e+01& 1.0e+02& 7.3e-02& 2.4e-01& 1.0e+00& 1.3e+00\\
   G22.7 & 1.4e+02& 1.6e+02& 4.1e-01& 4.5e-01& 1.7e+00& 2.5e+00\\
   3C396 & 3.0e+01& 8.8e+01& 3.6e-01& 5.8e-01& 2.0e+00& 3.0e+00\\
   G54.4 & 2.3e+01& 4.0e+01& 3.1e-02& 6.4e-02& 3.0e-01& 3.9e-01\\
  Kes 17 & 2.7e+01& 3.5e+01& 1.4e-01& 4.0e-01& 7.6e-01& 1.3e+00\\
  G311.5 & 2.3e+01& 1.1e+01& 1.5e-02& 1.9e-01& 1.8e+00& 2.0e+00\\
 RCW 103 & 1.1e+02& 1.3e+02& 8.4e-02& 5.1e-01& 5.0e-01& 1.1e+00\\
  G344.7 & 4.6e+01& 0.0e+00& 1.2e-02& 5.7e-07& 5.7e-01& 5.8e-01\\
  G348.5 & 3.7e+00& 1.1e+01& 2.1e-02& 2.0e-01& 4.3e-01& 6.4e-01\\
  CTB 37 & 2.3e+02& 2.8e+02& 1.1e+00& 3.4e+00& 1.1e+01& 1.5e+01\\
  G349.7 & 6.8e+00& 2.1e+00& 1.3e+00& 1.2e+01& 2.0e+01& 3.3e+01\\
\end{tabular}
\label{int_dust}
\end{table}

The correction factors are very large for many of the SNRs.  
Often time only
a few percent or less of the total dust emission was probed by the MIPS
SED.  The derived dust masses are therefore only to be considered
estimates. Further, it is clear from the derived correction factors between
the 24 and 70 $\mu$m images that some uncertainty is inherited here as
well.  There are several potential reasons for the different correction
factors.  The most likely is non-linearity effects in the 70 $\mu$m data
that will tend to underestimate the flux, in particular from brighter
emission spots. We estimate the masses determined to be good within a
factor of two to three. Nevertheless, the calculated dust masses are useful
to estimate the amount of dust in molecular clouds that is being processed
by SNRs.  Each SNR appears to be able to heat, and thus shatter, one to
several solar masses of dust at any time.  Due to the speed of the shock
front, it is likely it will penetrate further into the surrounding
molecular cloud and further process the dust.  Thus, the total amount of
dust that can be heated and processed by a single SNR can be up to ten
solar masses. 

One of the few previous estimates of ISM dust being processed by a SNR was
by \citet{reachrho}. They estimated the dust mass within one ISO LWS beam
(80\arcsec\ FWHM) in 3C391 to be 1 M$_\odot$. Integrating for the whole SNR
would increase this estimate substantially, possible up to 10 M$_\odot$.
However, the two other SNRs analyzed in the same paper, W28, and W44, had
an order of magnitude less dust within the beam. Further two of the SNRs
had previous dust mass estimates based on IRAS photometry \citep{saken92}.
They found a total amount of warm dust of 5.6$\times 10^{-3}$ M$_\odot$ for
Kes 17, and 0.17 M$_\odot$ for G349.7+0.2. Whereas the latter is in good
agreement with the mass estimated here, we derive a substantially larger
mass for Kes 17. The lower mass derived by \citet{saken92} may be due to
beam dilution and the limited sensitivity that would leave most of the
emission undetected with IRAS. 
 \citet{reachrho} estimated the mass within one beam of the ISO LWS camera to be roughly 1 M$_\odot$, comparable to a typical dust mass found here of 2-3 M$_\odot$.


We can estimate the time scale for processing an amount of dust comparable to the content in the Galaxy. 
A rough volume of the Galaxy can be estimated by assuming it is a cylinder with a height of 100 pc and a radius of 15 kpc. 
A typical SNR expands to some 20 pc and the rate of type II SNRs is roughly $1.9/100 ~yr$ \citep{diehl}. 
Thus, the time scale is then $\pi \times(15000)^2 \times 100/(4~\pi/3 \times 20^3/(1.9/100) ~yr\sim 110 ~Myr$ for Galactic dust to, on average,  have been processed once by a SNR shock.

\section{Conclusions} 
We have presented the results from MIPS SED and IRS low resolution spectroscopy of 14 Galactic SNRs. 
Our main results are the following; 
1) We find that they show sign for interactions with a surrounding molecular cloud. 
This is established through the presence of $H_2$ lines and the detection of [O~I] at 63 $\mu$m. This almost doubles the sample of known interacting SNRs. 
2) The dust continuum varies markedly from SNR to SNR, with relatively large variations in the relative dust species abundances. 
3) We find evidence for dust processing by comparing the dust species ratios. 
4) The integrated dust mass is estimated for a sub--set of the SNRs. 
We find typically around 1 M$_\odot$ of dust  is shocked by a SNR interacting with a molecular cloud. 
5) We find that the main cooling is occurring through the dust with relatively minor contribution from $H_2$ and [O~I]. 

\acknowledgements 
We thank the anonymous referee for careful reading and  insightful
comments which help to improve the paper. 
This work is based on observations made with the {\it Spitzer
Space Telescope}, which is operated by the Jet Propulsion Laboratory,
California  Institute of Technology, under NASA contract 1407. Partial
Support for this work was provided by both a NASA {\it Spitezr} GO award issued
by JPL/Caltech and a LTSA grant NRA-01-01-LTSA-013.
The DUSTEM code is available from http://www.ias.u-psud.fr/DUSTEM/

\end{document}